\documentclass{article}

\usepackage{microtype}
\usepackage{graphicx}
\usepackage{booktabs} 

\usepackage{array}
\usepackage{xspace}
\usepackage{amsmath}
\usepackage{subcaption}
\usepackage{enumitem}
\usepackage{amssymb}
\usepackage{multirow} 

\usepackage{hyperref}

\usepackage[accepted]{mlsys2025}

\newcolumntype{M}[1]{>{\centering\arraybackslash}p{#1}}
\newcommand{\sysname}{Astra\xspace}
\newcommand{\sssec}[1]{\vspace*{0.05in}\noindent\textbf{#1}}

\mlsystitlerunning{
\sysname: Efficient and Money-saving Parallel Strategies Search on Heterogeneous GPUs
}

\begin{document}

\twocolumn[
\mlsystitle{\sysname: Efficient and Money-saving Automatic Parallel Strategies Search on Heterogeneous GPUs}



\mlsyssetsymbol{equal}{*}

\begin{mlsysauthorlist}
\mlsysauthor{Peiran Wang}{equal,baidu}
\mlsysauthor{Haibing Li}{equal,baidu}
\mlsysauthor{Haohan Fu}{baidu}
\mlsysauthor{Shiyong Li}{baidu}
\mlsysauthor{Yanpeng Wang}{baidu}
\mlsysauthor{Dou Shen}{baidu}
\end{mlsysauthorlist}

\mlsysaffiliation{baidu}{Baidu, inc., Beijing, China}


\mlsyskeywords{Machine learning system, parallel computing, cloud computing}

\vskip 0.3in

\begin{abstract}
In this paper, we introduce an efficient and money-saving automatic parallel strategies search framework on both homogeneous and heterogeneous GPUs: \sysname.
First, \sysname searches for the efficiency-optimal parallel strategy in both GPU configurations search space and parallel parameters search space.
Then, \sysname also provides the solution on heterogeneous GPUs by mathematically modeling the time consumption of heterogeneous training.
At last, \sysname is the first to propose the automatic parallel strategy search on money-saving.
The experiment results demonstrate that \sysname can achieve better throughput than expert-designed strategies.
The search time cost for \sysname can also be limited to 1.27 seconds in a single-GPU setting and less than 1.35 minutes in a heterogeneous-GPU setting on average with an accuracy of over 95\%.
\end{abstract}
]




\section{Introduction}

\begin{figure}[htb!]
\centering
\includegraphics[width=0.48\textwidth]{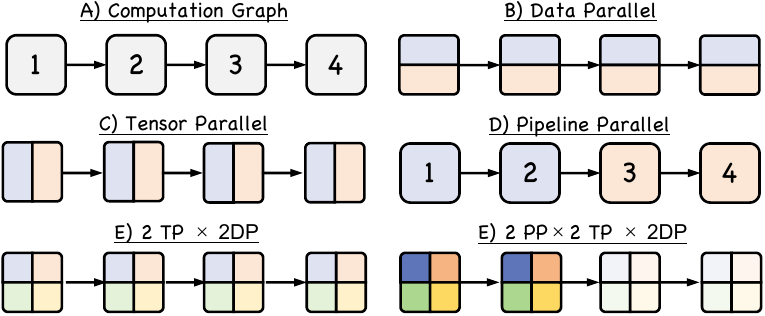}
\caption{
Different parallel methods have been proposed: b) tensor parallelism, c) data parallelism, d) pipeline parallelism, etc.
Furthermore, a new parallelism paradigm, hybrid parallelism, has been primarily applied in real-world applications that combines current parallelism methods.
}
\label{fig:parallel}
\end{figure}

The large language models like GPT-4 \cite{achiam2023gpt}, Llama \cite{touvron2023llama}, Llama2 \cite{touvron2023llama2}, Mistral \cite{jiang2023mistral}, etc., are consisted of billions of parameters and has achieved remarkable capabilities in natural language processing and other domains.
Training such large models requires the parallel computing of numerous GPUs, and numerous GPU parallel training methods have emerged, such as data parallel, tensor parallel, pipeline parallel, etc.
These methods have different advantages and disadvantages, for instance, tensor parallel saves memory consumption but raises the training latency.
There is not a single parallel method that is better than other parallel methods at each scene.

Thus, in real-world practice, LLM developers mainly combine those parallel methods into a hybrid parallel strategy to satisfy their needs.
Many industry developer teams rely on human experts to manually design hybrid parallel strategies based on their experience.
Other than relying on manual expert-designed hybrid parallel strategies, previous researchers have also proposed \textbf{automatical parallelization}, such as Alpa\cite{zheng2022alpa}, Aceso \cite{liu2024aceso} and Galvatron\cite{miao2022galvatron}, which automatically searches throughput-optimal hybrid parallel strategies based on cost models and optimization algorithms. 

However, previous methods still leave several realistic and important questions to answer.
Now, assume you are a developer of a GPU training cloud service:

\sssec{RQ 1-Large search space}: The GPU training cloud platform will have many customers with different needs.
Different customers have different requirements for model type, model scale, GPU scale (tens of cards to tens of thousands of cards), and GPU type.
Most customers of the GPU training cloud platform are novice users (otherwise they would build their private cloud).
When designing training strategies, they need a fast, accurate, and low-cost tool to obtain the best GPU configuration (GPU number) and the corresponding splitting parameters to guide training.
However, previous methods only focus on the scenes of fixed GPU numbers and GPU types.

\sssec{RQ 2-Heterogeneous GPUs}:
In pursuit of extreme cost, many customers of cloud platform providers purchase different types of GPUs for heterogeneous training.
The current method focuses on the scenario of a single GPU type and cannot solve the parallel strategy generation in this heterogeneous GPU heterogeneous training scenario.

\sssec{RQ 3-Money saving}:
In addition to focusing on training efficiency, cloud providers' customers may also be sensitive to the cost of model training.
For example, when a startup wants to train a large model for business, it often wants a more cost-effective training strategy due to initial funding constraints, rather than a training strategy with the best training efficiency.
The current solution only supports searching for the strategy with the best training efficiency.

To solve the above problems, we designed and implemented \sysname. \sysname supports three search modes.
First, homogeneous mode, specify a single GPU type and GPU number, \sysname can search for the optimal strategy.
Second, heterogeneous mode, in the heterogeneous training scenario of mixing different types of GPUs, we are the first to completely solve the simulation of a heterogeneous-GPU training scene, and build mathematical modeling and theoretical problems in heterogeneous-GPU scenarios, which can also search for the optimal strategy of heterogeneous GPUs.
Third, cost mode, with specifying the maximum number of GPUs and maximum money limit, \sysname also supports searching for the most efficient one.
Finally, the performance simulation efficiency of \sysname is very fast, which is 98.7\% lower than that of industry competitors. The performance simulation accuracy of \sysname is more than 95\%, which is better than expert strategies in most scenarios. \sysname has a low computational cost, it can run locally without high computing power need, and can search for the best strategy in 1.27 seconds for a single-GPU setting and less than 1.35 minutes for a heterogeneous-GPU setting on average with an accuracy of over 95\%.

\section{Related Work}\label{sec:moti}

In this section, we discuss the related works of \sysname.
First, we first discuss current large-scale distributed deep learning model training in \S\ref{sec:moti:large}.
Then, we introduce different parallel methods and the large search space of hybrid parallel in \S\ref{sec:moti:parallel}, which drives the need to build auto parallel for large language model training.

\subsection{Large-scale Distributed DL Training}\label{sec:moti:large}

\par In recent years, large-scale Transformer models have demonstrated exceptional performance across various application domains, pushing the frontiers of deep learning research. For instance, models like BERT \cite{devlin2018bert}, GPT-3 \cite{achiam2023gpt}, and T5 \cite{ni2021sentence} have achieved state-of-the-art results in natural language processing tasks such as machine translation, text classification, and question answering. These advancements underscore the capabilities of large-scale models in capturing complex patterns and generating high-quality outputs that surpass those of smaller models.

\par However, the sheer size of these models poses significant computational challenges. GPT-3, for example, consists of 175 billion parameters \cite{achiam2023gpt}, making it impractical to train on a single GPU due to memory and processing constraints \cite{hoffmann2022empirical}. Single-GPU setups cannot handle the extensive computations and data storage required by such massive models \cite{hoffmann2022empirical}. As model sizes increase, the demand for computational power and memory scales exponentially, further exacerbating the limitations of single-GPU environments \cite{geiping2023cramming}.

\par To address these challenges, the training of large-scale Transformer models necessitates the use of distributed training across multiple GPUs \cite{narayanan2021efficient, li2020pytorch, pipedream}. This approach leverages several GPUs' combined computational power and memory capacity, enabling the efficient handling of large models. 


\subsection{Different Parallel Methods}\label{sec:moti:parallel}

\par Existing parallel methods for training large-scale models include tensor parallelism \cite{narayanan2021efficient}, data parallelism \cite{hillis1986data, li2020pytorch, agarwal2012reoptimizing}, and pipeline parallelism \cite{li2021terapipe, kim2023bpipe, yang2021pipemare, narayanan2021memory, zhao2021v}, each offering distinct advantages and catering to different aspects of the training process (see Fig. \ref{fig:parallel}). 

\sssec{Tensor parallelism} involves partitioning the model's tensors (such as weight matrices) across multiple GPUs, allowing parallel computation of operations like matrix multiplications. 
This approach effectively reduces the memory footprint on each GPU by distributing the model parameters and computations, thereby accommodating larger models that would otherwise exceed the memory capacity of a single GPU \cite{narayanan2021efficient}.

\sssec{Data parallelism}, on the other hand, splits the training data across multiple GPUs \cite{hillis1986data, li2020pytorch, agarwal2012reoptimizing}. Each GPU maintains a complete copy of the model and processes a subset of the data. The gradients calculated by each GPU are then averaged and synchronized across all GPUs to ensure consistent model updates. This method is particularly effective in leveraging the computational power of multiple GPUs to handle large datasets and improve training speed. 

\sssec{Pipeline parallelism} divides the model into sequential stages, each assigned to a different GPU \cite{li2021terapipe, kim2023bpipe, yang2021pipemare, narayanan2021memory, zhao2021v}. As data flows through the pipeline, each GPU processes its assigned stage sequentially. This approach enables efficient utilization of GPU resources by maintaining continuous data flow and minimizing idle times, especially for intense models with numerous layers.

\sssec{Manual hybrid parallelism}. In practice, experts often combine these parallelism techniques to create hybrid parallelism strategies tailored to the specific requirements of their models and hardware configurations \cite{song2019hypar}. 
This expert-crafted hybrid parallelism involves significant manual effort and domain knowledge to balance the trade-offs between computation, memory usage, and communication overheads. 
For instance, systems like DeepSpeed \cite{rasley2020deepspeed} and Megatron-LM \cite{narayanan2021efficient} employ a mix of data parallelism, tensor parallelism, and pipeline parallelism to optimize the training of massive models like GPT-3 \cite{achiam2023gpt}.

\sssec{Automatic parallelism}. However, manual hybrid parallelism has its limitations. The complexity and diversity of modern deep learning models and hardware environments make it increasingly difficult for experts to design optimal parallelism strategies. 
This is where automatic parallelism comes into play. Automatic parallelism systems, such as Alpa \cite{zheng2022alpa}, Metis \cite{um2024metis}, and Galvatron \cite{miao2022galvatron}, automate finding efficient parallelism plans by exploring a vast search space of potential strategies using advanced algorithms and heuristics.

\section{Proposed Scheme}\label{sec:scheme}


\begin{figure*}[htb!]
\centering
\includegraphics[width=\textwidth]{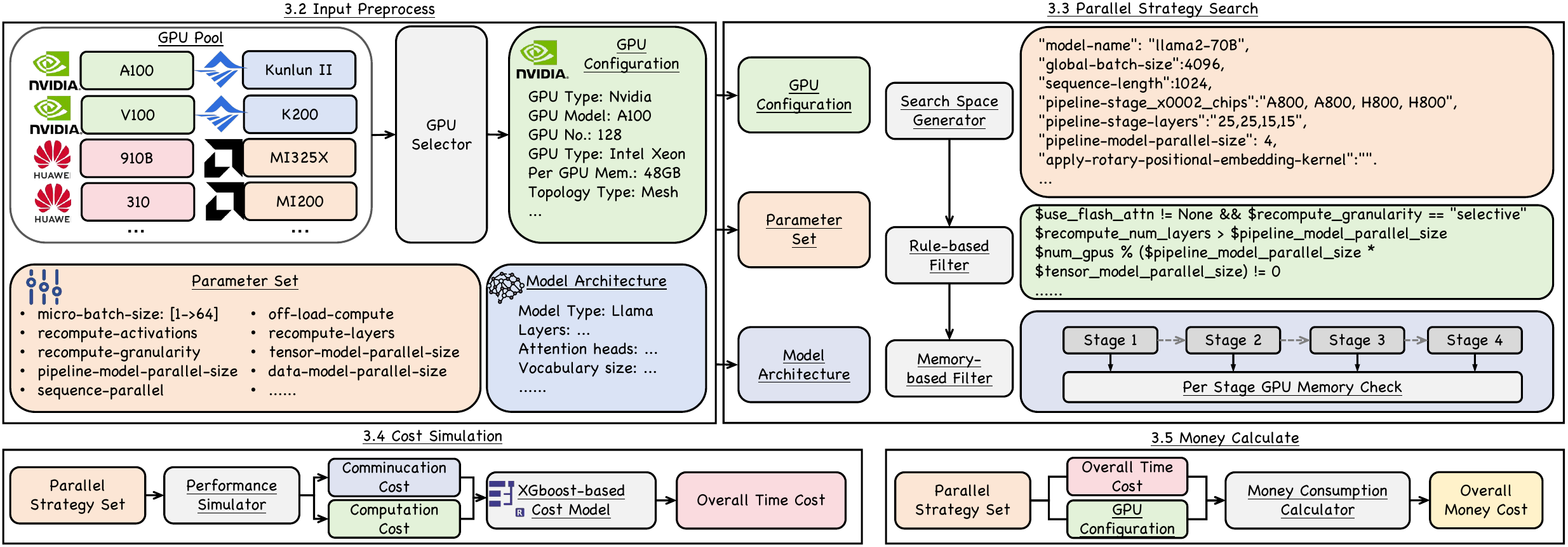}
\caption{
\sysname works as follows:
1) \underline{Input Preprocess}: \sysname extracts MegatronLM's parameter set as its parameter search space, parses the model architecture, and generates diverse GPU configurations based on GPU type, model, and quantity.
2) \underline{Parallel Strategy Search}: Using GPU configurations, parameter set, and model architecture, \sysname creates parallel strategies. User-defined rules and memory constraints filter these strategies.
Next, the memory-based filter computes per-stage's allocated GPU's memory. The strategy is filtered if the memory is out of the upper boundary.
3) \underline{Cost Simulation}: The simulator calculates communication and computation costs using an XGBoost model to estimate each operator's time, determining the total time for each strategy.
4) \underline{Money Calculation}: It computes the monetary cost of each strategy based on time and GPU configurations.
}
\label{fig:scheme}
\vspace{-15pt}
\end{figure*}


\sysname uses MegatronLM as the runtime backend due to various advantages of it (\S\ref{sec:scheme:runtime}).
\sysname works as follows:

1) \underline{Input Preprocess} (\S\ref{sec:scheme:input}): \sysname first extracts the parameter set of MegatronLM as the parameter search space of \sysname. Then, the training model architecture is parsed. Next, the GPU selector of \sysname iterates the GPU pool to generate a diverse set of GPU configurations, including the GPU type, model, number, etc.

2) \underline{Parallel Strategy Search} (\S\ref{sec:scheme:search}): Based on the three input information (GPU configurations, parameter set, model architecture), \sysname runs its search space generator to generate a diverse set of parallel strategies.
Then, the rule-based filter reads users' crafted rules to filter out the strategies not permitted by user-crafted rules.
Next, the memory-based filter computes the memory allocated to each stage's GPU. If the memory is outside the upper boundary, the strategy is filtered.

3) \underline{Heterogeneous GPU Search} (\S\ref{sec:scheme:heter}):
We re-modeled the time cost of heterogeneous training so that \sysname can search for heterogeneous training strategies. We also performed a computational complexity analysis on the search overhead of heterogeneous training.

3) \underline{Cost Simulation} (\S\ref{sec:scheme:cost}): Then, given the filtered parallel strategies, the performance simulator computes the communication cost and computation cost to get the overall time cost for each parallel strategy.
Each operator's time cost is predicted by a trained XGBoost model.

4) \underline{Money Calculation} (\S\ref{sec:scheme:money}): At last, \sysname calculates each parallel strategy's money cost for the training models.
The money computation calculator calculates the money cost for each parallel strategy due to the overall time cost and GPU configurations.

\subsection{MegatronLM-based Runtime}\label{sec:scheme:runtime}

\sysname leverages MegatronLM as the runtime backend for the parallel strategy execution.
Below, we describe MegatronLM's key features and benefits, substantiating its selection as our system's core runtime.

\sssec{Overview of MegatronLM}. MegatronLM is a highly optimized, scalable framework for training large-scale Transformer models. Developed by NVIDIA, it has been widely adopted in research and industry for its robust performance and flexibility in handling diverse parallelism techniques. 
We utilize MegatronLM as the \sysname's backend for the reasons below:

\begin{enumerate}[label={[\arabic*]}, itemsep=0pt, leftmargin=*,topsep=0pt]
    \item \textbf{Support for diverse parallelism methods}. MegatronLM is designed to handle a variety of parallelism strategies, including data parallelism, tensor parallelism, and pipeline parallelism. This versatility in parallelism methods allows MegatronLM to optimize training for different model architectures and hardware configurations, ensuring maximum efficiency.
    \item \textbf{Industry-validated reliability}. MegatronLM's reliability has been extensively validated in industrial applications. It is widely used in the industry for training some of the largest language models, such as GPT-3, demonstrating its robustness and stability. Furthermore, as an open-source framework, continuous contributions from both the research community and industry practitioners ensure that MegatronLM stays up-to-date with the latest advancements and best practices in deep learning.
\end{enumerate}




\subsection{Input Preprocess}\label{sec:scheme:input}

The input preprocess stage in \sysname\ involves several key steps to prepare and organize the data necessary for the subsequent parallel strategy search. This stage ensures the system can effectively explore various parallelization strategies and configurations to optimize model training. Here’s a detailed breakdown of the process:

\sssec{GPU pool}. 
\sysname supports three search modes, and generates the corresponding GPU pool:

\underline{Mode 1: homogeneous}: Specify a GPU type and the number of cluster GPUs, such as specifying A800 and 32768 GPUs, to find the best strategy:
\begin{equation}
    \mathcal{C}_{gpu}=(A800, 32768)
\end{equation}
\underline{Mode 2: heterogeneous}: Specify multiple GPU types and the number of cluster GPUs, such as specifying the number of cluster GPUs to be 8192, and the cluster uses two types of GPUs, A800 and H100 (the maximum number of these two GPUs can be specified at the same time), to find the best heterogeneous training strategy:
\begin{equation}
    \mathcal{C}_{gpu}=8192, (A800:2048),(H100:7168)
\end{equation}
where maximum number of A800 is 2048, maximum number of H100 is 7168.

\underline{Mode 3: cost}: Specify a GPU type(for example, H100) the maximum number of cluster GPUs(for example, 4096) and the maximum money limit to find the best strategy:
\begin{align}
\mathcal{C}_{gpu}=\{
[(H100, 2)], [(H100, 4)], ... [(H100, 4096)]\}
\end{align}
Each configurations defines a runnable GPU collections for \sysname.

\sssec{Parameter set extraction}. \sysname\ extracts a detailed parameter set $\mathcal{P}$ from the MegatronLM framework, including micro-batch size, recompute activations, pipeline model parallel size, sequence parallel, tensor model parallel size, data model parallel size, off-load compute, recompute layers, etc. These parameters define the search space for possible parallelization parameter set:
\begin{equation}
    \mathcal{P} = \left\{ p_1, p_2, \dots, p_n \right\},
\end{equation}

\sssec{Model architecture parsing}. The system parses the architecture of the training model, denoted as $\mathcal{M}$. This includes details such as:
\begin{align}
    \mathcal{M} = \{ \text{model type}, \text{number of layers},
    \text{hidden size},
    \\\text{attention heads},
    \text{intermediate size},
    \text{vocabulary size}\}
\end{align}

\sssec{Integration of input information}. The selected GPU configurations $\mathcal{C}_{gpu}$, parameter set $\mathcal{P}$, and parsed model architecture $\mathcal{M}$ are integrated to form the foundational input for the parallel strategy search:
\begin{equation}
    \mathcal{I} = (\mathcal{C}_{gpu}, \mathcal{P}, \mathcal{M})
\end{equation}

\subsection{Parallel Strategies Search}\label{sec:scheme:search}

The parallel strategy search stage generates, filters, and identifies the most efficient parallelization strategies for training large-scale Transformer models. This stage consists of three components: 
(1) The \textbf{search space generator} generates diverse parallel strategies, 
(2) the \textbf{rule-based filter} filters illegal parallel strategies according to user-defined rules, and 
(3) the \textbf{memory-based filter} filters out parallel strategies that are out of memory bounds.

\sssec{Search space generator}. 
The search space generator iterates all GPU configurations $\mathcal{C}_{gpu}$ according to the user input,
all potential parallel parameters value denoted as $f(\mathcal{P})$ with the fixed model architecture $\mathcal{M}$ to get all potential $s_i \in \mathcal{S}$.
Each strategy is defined as
\begin{equation}
    s_i=\{ c_{gpu}, P', \mathcal{M} \}, c_{gpu} \in \mathcal{C}_{gpu}, P' \in f(\mathcal{P})
\end{equation}

The total number of strategies $\mathcal{S}$ is determined by the product of all parameter options and the number of all GPU configurations:
\begin{equation}
    |\mathcal{S}| = \prod_{P' \in f(\mathcal{P})} |P'| \times |\mathcal{C}_{gpu}|
\end{equation}

\sssec{Rule-based filter}. The Rule-Based Filter applies user-defined rules to filter out strategies that meet specific rules. Let $r_1, r_2, \dots, r_k$ represent the rules. A strategy $s_i \in \mathcal{S}$ is valid if:
\begin{equation}
    r_j(s_i) = \text{False} \quad \forall j \in \{1, 2, \dots, k\}
\end{equation}
which suggests that if the strategy meets any of the rules, it is dropped.

Each rule is a logical expression involving the strategy parameters, such as:
\begin{align}
    r_j = (\$use\_flash\_attn \neq \text{None}) \land 
    \\
    (\$recompute\_granularity = \text{selective})
\end{align}
If any rule is met, the strategy is filtered out.

To illustrate the design of our rule-based filters, we provide three examples of rules:

1. Flash attention rule:
\begin{align}
    \$use\_flash\_attn \neq \text{None} \quad \&\& 
    \\
    \quad \$recompute\_granularity = \text{selective}
\end{align}
This rule ensures that the flash attention is being used, 
and the recompute granularity can not be selective.

2. Layer recomputation rule:
\begin{align}
    \$recompute\_num\_layers > 
    \\
    \$pipeline\_model\_parallel\_size
\end{align}
This rule filters out the strategies where the number of layers to be recomputed exceeds the number of pipeline parallel stages.

3. GPU division rule:
\begin{align}
    \$num\_gpus \% (\$pipeline\_model\_parallel\_size \times 
    \\
    \$tensor\_model\_parallel\_size) \neq 0
\end{align}
This rule ensures that the total number of GPUs is divisible by the product of the pipeline model parallel size and tensor model parallel size.

These rules are written in a format:
\begin{equation}
    \text{expression} \ \&\&/\| \ \text{expression} \ \&\&/\| \ \text{expression} \dots
\end{equation}
where $\&\&$ has a higher precedence than $\|$, and the expressions are evaluated from left to right.

\sssec{Memory-based filter}. The Memory-Based Filter calculates the memory usage for each stage of a strategy. Let $M_i(s_j)$ denote the memory required for stage $i$ of strategy $s_j$. The strategy is filtered out if:
\begin{equation}
    M_i(s_j) > M_{gpu}, \quad \forall i, s_j
\end{equation}

Only strategies that satisfy the memory constraints for all stages are retained for further evaluation:
\begin{equation}
    \mathcal{S}_{valid} = \left\{ s_j \in \mathcal{S} \mid M_i(s_j) \leq M_{gpu}, \forall i \right\}
\end{equation}

Specifically, by conducting extensive offline experiments, we collect statistics on the memory consumption of a single layer under various configurations (e.g., enabling/disabling flash attention, enabling/disabling selective recomputation, enabling/disabling sequence parallelism, and simultaneously enabling both sequence parallelism and flash attention). Through these experiments, we derive an empirical formula for single-layer memory usage as a function of parameters such as the number of microbatches $b$, sequence length, hidden size, FFN size, tensor parallelism (TP), pipeline parallelism (PP), and the number of attention heads. This empirical formula allows us to estimate the total memory consumption for any given strategy based on the configuration settings.

\begin{figure*}[htb!]
\centering
\includegraphics[width=\textwidth]{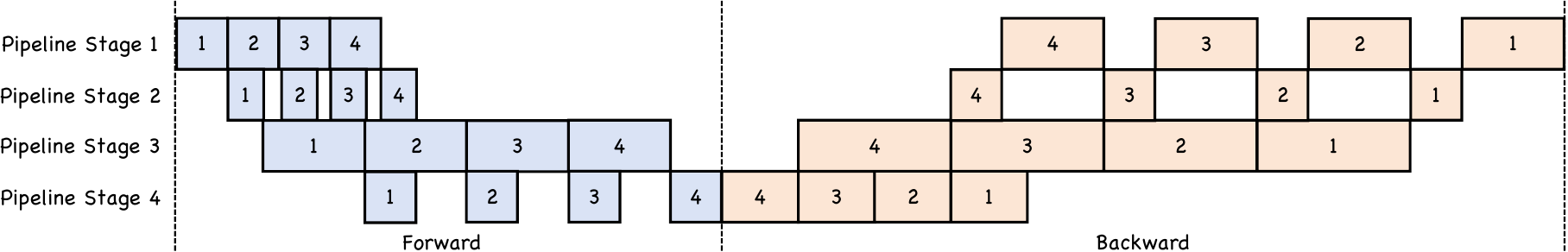}
\caption{
    The time cost of each pipeline stage is different, and the bubble time is also different, so the total duration cannot be converted by the duration of a pipeline stage and the bubble time.
}
\label{fig:heter}
\end{figure*}

\subsection{Heterogeneous GPUs Strategies Search}\label{sec:scheme:heter}

Compared with homogeneous GPU strategies search, heterogeneous GPU strategies search has several problems:
\begin{enumerate}[label={[\arabic*]}, itemsep=0pt, leftmargin=*,topsep=0pt]
    \item In a heterogeneous GPU scene, the time cost of each pipeline stage varies, and each pipeline's corresponding bubble time also differs. The classic total duration formula: 
    $
    T_{total} = T_{comp} + T_{comm} + T_{bubble}
    $
    with
    $
    T_{bubble}=\frac{pp - 1}{m} \times (T_{comp}+T_{comm})
    $
    is no longer applicable.
    \item Each pipeline stage can be a different GPU type. The pipeline stages of different GPUs may have different numbers of model layers. The number of such combinations increases exponentially with the GPU type and the model number layers. 
\end{enumerate}

\sssec{Definition}. Assume there are $M$ types of GPUs, with the computational capability of the $i$-th type denoted as $c_i$, and a maximum quantity of $l_i$ for each type. Set the following parameters: pipeline parallel size = $P$, data parallel = $D$, tensor parallel = $T$, number of micro-batches = $K$, model layers = $N$.

Consider deploying $P$ pipeline stages across $M$ types of GPUs. The partition of $P$ stages across $M$ types of GPUs is denoted by $\{p_1, p_2, \dots, p_P\}$, where each $p_i$ is a value representing the GPU type it is deployed on, ranging from $1$ to $M$. Thus, there are $O(M^P)$ possible configurations. An example of a simple pipeline parallelization plan (assuming four pipeline stages deployed on four types of GPUs, each stage containing a number of model layers $n_1, n_2, n_3, n_4$) is shown in Fig. \ref{fig:heter}. Each pipeline stage has a different computation latency, and the bubble time varies across stages, so the total latency cannot simply be the sum of the latencies and bubble times across all pipeline stages. 

\sssec{Heterogeneous GPU cost model}. Consider any given partition where the $P$ pipeline stages are divided across $M$ types of GPUs. Let $t_{p_i}$ represent the computation latency of the $i$-th pipeline stage, and $h_{p_i}$ represent the peer-to-peer communication latency. The forward latency for this partition (the backward latency is similar) can be derived as follows:

\begin{equation}
\label{eq:heter:def}
    \sum_{1 \leq i \leq P} (t_{p_i} + h_{p_i}) + (K - 1) \times \max(t_{p_i} + h_{p_i})
\end{equation}

Here, $t_{p_i}$ depends only on the number of model layers in the $i$-th pipeline stage and the GPU it is deployed on, while $h_{p_i}$ depends solely on the tensor shape during communication. Therefore, for any configuration, the order of $p_i$ can be rearranged to place identical values of $p_i$ in consecutive positions (i.e., each segment of $P$ pipeline stages is deployed on the same GPU type), reducing the total number of partitions from $O(M^P)$ to $C_{P-1}^{M-1} \times (M-1)! \approx O(P^{M-1})$.

In summary, when the model is deployed in a cluster with $M$ types of GPUs, the mathematical model of its strategy search space is equivalent to solving the following equation for $m_i$, $n_i$, and $n_i$ (assuming the number of model layers in the pipeline stage on GPU type $i$ is $n_i$):

\begin{equation}
\label{eq:heter:comp}
\left\{
\begin{array}{l}
m_i, n_i \left| \sum_{1 \leq i \leq M} m_i = P, \quad m_i \leq \frac{l_i}{D \cdot T}, \right. \\
\left. \quad \sum_{1 \leq i \leq M} m_i \cdot n_i = N \right.
\end{array}
\right\}
\end{equation}

\sssec{Search complexity analysis}. Dividing $P$ pipeline stages into $M$ sequential segments results in $\binom{P-1}{M-1} \times (M-1)! \approx O(P^{M-1})$ possible configurations. Solving Equation (\ref{eq:heter:comp}) for all combinations of $m_i$ has a time complexity of $O(P^{M-1})$, while for any fixed combination of $m_i$, solving Equation (\ref{eq:heter:comp}) for all $n_i$ has a time complexity of $O\left(\frac{N}{m_1} \times \frac{N}{m_2} \times \dots \times \frac{N}{m_{M-1}}\right) < O(N^{M-1})$.

Thus, the total time complexity for all combinations of $m_i$ and $n_i$ for Equation (\ref{eq:heter:comp}) is $O(N^{M-1}) \times O(P^{M-1})$.

\subsection{Cost Simulation}\label{sec:scheme:cost}

The Cost Simulation stage is essential for evaluating the performance of the filtered parallel strategies from the Parallel Strategy Search stage. This stage estimates the overall training time by considering communication and computation costs.

\sssec{Input: parallel strategy set}. The parallel strategies that pass through the Rule-Based Filter and Memory-Based Filter, denoted as $\mathcal{S}_{valid}$, are considered for cost simulation. These strategies represent feasible configurations that satisfy all the constraints and are now evaluated for their computational and communication efficiency. Let $\mathcal{S}_{cost}$ represent the set of strategies selected for cost evaluation:
\begin{equation}
    \mathcal{S}_{cost} = \{ s_1, s_2, \dots, s_m \}, \quad s_i \in \mathcal{S}_{valid}
\end{equation}

\begin{figure}[htb!]
\centering
\includegraphics[width=0.48\textwidth]{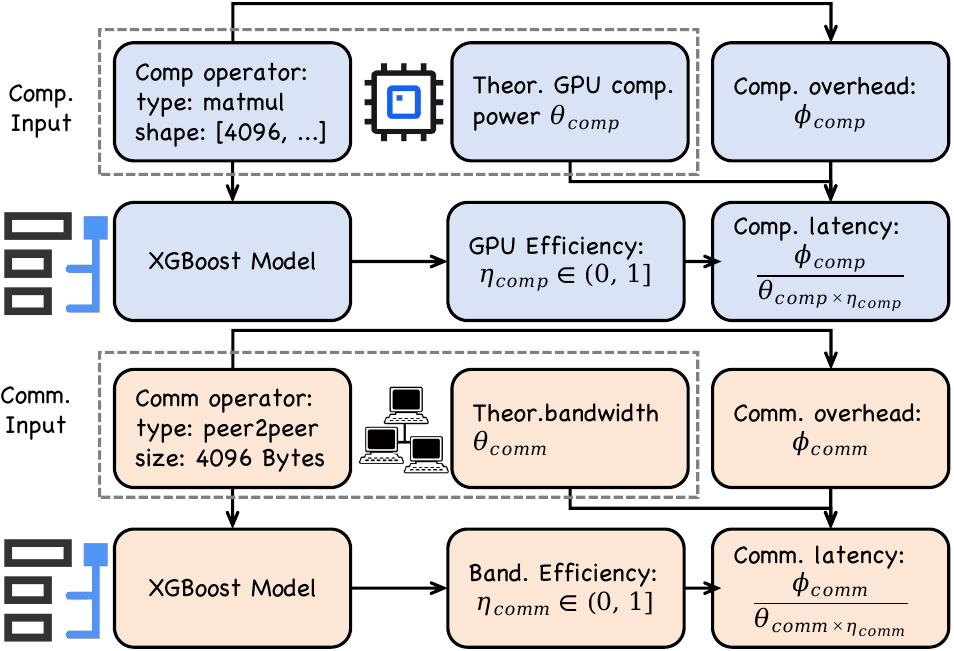}
\caption{
\sysname' cost model is based on the XGBoost model.
Each computation and communication operator's latency cost is calculated based on XGBoost's prediction on efficiency and theoretical computing power and communication bandwidth.
}
\label{fig:xgboost}
\end{figure}

\sssec{XGBoost-based cost model}. The computation time for each operator is estimated as follows (see Fig. \ref{fig:xgboost}):

\begin{equation}
T_{comp}(s_i) = \frac{\theta_{comp}}{ \phi_{comp} \times \eta_{comp}},
\end{equation}
where $\theta_{comp}$ is the theoretical computing overhead, $\phi_{comp}$ is the theoretical computing power, $\eta_{comp}$ represents the GPU efficiency, a parameter estimated via XGBoost, with values constrained within the range $(0, 1]$.

Similarly, the communication time for each operation is calculated as:

\begin{equation}
T_{comm}(s_i) = \frac{\theta_{comm}}{ \phi_{comm} \times \eta_{comm}},
\end{equation}
where $\theta_{comm}$ is the theoretical communication overhead, $\phi_{comm}$ is the theoretical bandwidth, $\eta_{comp}$ represents the bandwidth efficiency, a parameter estimated via XGBoost, with values constrained within the range $(0, 1]$.

One of the distinguishing features of the \sysname framework compared to other partitioning tools lies in its unique approach to estimating operator latency. 
Unlike conventional methods that rely on a database of pre-measured values, \sysname calculates the operation latency analytically, which enables automatic adaptation to new models and operators. 
This capability to dynamically adjust to diverse computational requirements is a core reason behind \sysname’s adaptability to novel architectures and computational operators.






\sssec{Performance simulator}. 
The Performance Simulator first computes the time cost for each pipeline stage $p_i \in \mathcal{S}_{cost}$ by evaluating two key components: 
computation cost $T_{comp}(p_i)$, communication cost $T_{comm}(p_i)$
\begin{align}
    T_{total}(p_i) &= \sum^{op_{comp}\in p_i}_{op_{comp}}T_{comp}(p_i, op_{comp}) 
    \\
    &+ \sum^{op_{comm}\in p_i}_{op_{comm}}T_{comm}(p_i, op_{comm})
\end{align}
And then, performance simulator utilizes the equation (\ref{eq:heter:def}) to compute the overall performance in both heterogeneous and homogeneous setting.

\subsection{Money-limit Search}\label{sec:scheme:money}

In this section, we present the process for calculating the monetary cost of parallel strategies in the \sysname framework. The method involves traversing all possible strategies, selecting the optimal strategies based on throughput and cost, and finally computing the financial cost for each strategy. This approach builds upon the principles outlined in \textit{Cost-based \sysname Framework}, where the highest throughput strategy is first identified without considering cost limitations.

\sssec{Optimal strategy pool}. Once all possible valid strategies have been generated, we aim to create an optimal pool of strategies. Let $P_i$ represent the throughput of strategy $i$ and $C_i$ its corresponding cost. We define the optimal pool such that for any strategy $(P_i, C_i)$ in the pool, no other strategy $(P_j, C_j)$ exists such that:
\begin{equation}
    P_j > P_i \quad \text{and} \quad C_j < C_i
\end{equation}

The goal is to find the set of strategies $\mathcal{S}_{opt}$ where:
\begin{align}
    \mathcal{S}_{opt} = \{ (P_i, C_i) \mid \nexists (P_j, C_j) 
    \\
    \text{ such that } P_j > P_i \land C_j < C_i \}
\end{align}

\sssec{Money cost calculation}. The monetary cost for each strategy in the optimal pool is calculated based on the GPU usage and the total training time. Let $T_i$ be the total time required for strategy $i$, and let $g_i$ denote the cost per hour for using GPU $g$. The total cost $M_i$ for strategy $i$ is given by:
\begin{equation}
    M_i = T_i \times \mathcal{N}_{g_i} \times \mathcal{F}_{g_i}
\end{equation}
where $\mathcal{F}_{g_i}$ is the fee for GPU $g_i$ per second, $\mathcal{N}_{g_i}$  is the number of GPUs used by strategy.

Using this formula, we can compute the monetary cost for all strategies in the optimal pool. The strategy with the highest throughput  that meets the user's money constraints is selected as the final strategy.

\sssec{Sorting strategies by cost and throughput}. To sort the strategies, we first rank them by throughput in descending order. If two strategies have the same throughput, we then compare them by cost in ascending order. Formally, the sorting function $S$ can be expressed as:
\begin{equation}
    S(P, C) = \begin{cases} 
    P_i > P_j, & \text{if } P_i \neq P_j \\
    C_i < C_j, & \text{if } P_i = P_j 
    \end{cases}
\end{equation}

This ensures that the strategies are sorted optimally for both performance and cost.
\section{Implementation}

\sssec{Searchable Parameters}.
We implemented \sysname's searched parallel strategy on a MegatronLM backend.
Since we need to search out different parameters for different parallel strategies, we listed our parameter search space in Appendix Table \ref{tab:parameter}.

\sssec{Searchable GPUs}.
We implemented our system on Nvidia A800. Our hardware configurations enable each node to consist of 8 GPUs with each one connecting with Nvilink. Then for the cross-node GPU communication, we use PCIE.
\begin{table*}[t]
\centering
\tiny
\begin{tabular}{|M{1.2cm}|M{0.7cm}|M{1cm}|M{1cm}|M{1cm}|M{0.8cm}|M{1.2cm}|M{0.7cm}|M{1cm}|M{1cm}|M{1cm}|M{0.8cm}|}
\hline\hline
Model & \#GPU & \#Strategies & Search Time(/s) & Simulation Time(/s) & E2E Time(/s) & Model & \#GPU & \#Strategies & Search Time(/s) & Simulation Time(/s) & E2E Time(/s) \\ \hline
\multirow{4}{*}{Llama-2-7B} & 64 & 23348 & 0.06 & 49.7 & 51.0 & \multirow{4}{*}{Llama-2-13B} & 64 & 23400 & 0.05 & 58.1 & 59.3 \\ \cline{2-6} \cline{8-12} 
 & 256 & 14372 & 0.05 & 43.5 & 44.4 &  & 256 & 13552 & 0.03 & 49.9 & 50.8 \\ \cline{2-6} \cline{8-12} 
 & 1024 & 8856 & 0.04 & 41.8 & 42.2 &  & 1024 & 8920 & 0.02 & 51.0 & 51.7 \\ \cline{2-6} \cline{8-12} 
 & 4096 & 4700 & 0.03 & 33.0 & 33.2 &  & 4096 & 4720 & 0.02 & 44.1 & 44.3 \\ \hline
\multirow{4}{*}{Llama-2-70B} & 64 & 53264 & 0.1 & 68.8 & 75.0 & \multirow{4}{*}{Llama-3-8B} & 64 & 23348 & 0.05 & 48.3 & 49.6 \\ \cline{2-6} \cline{8-12} 
 & 256 & 31440 & 0.06 & 57.7 & 60.9 &  & 256 & 14372 & 0.04 & 42.0 & 42.8 \\ \cline{2-6} \cline{8-12} 
 & 1024 & 20152 & 0.05 & 57.4 & 59.6 &  & 1024 & 8856 & 0.03 & 40.9 & 41.3 \\ \cline{2-6} \cline{8-12} 
 & 4096 & 10948 & 0.04 & 63.2 & 65.0 &  & 4096 & 4700 & 0.03 & 32.7 & 32.9 \\ \hline
\multirow{4}{*}{Llama-3-70B} & 64 & 53264 & 0.1 & 66.8 & 71.8 & \multirow{4}{*}{GLM-67B} & 64 & 20528 & 0.04 & 19.3 & 20.6 \\ \cline{2-6} \cline{8-12} 
 & 256 & 31440 & 0.07 & 56.3 & 59.6 &  & 256 & 12132 & 0.03 & 16.6 & 17.4 \\ \cline{2-6} \cline{8-12} 
 & 1024 & 20152 & 0.05 & 55.5 & 57.6 &  & 1024 & 7948 & 0.02 & 16.9 & 17.3 \\ \cline{2-6} \cline{8-12} 
 & 4096 & 10948 & 0.04 & 62.4 & 63.4 &  & 4096 & 4196 & 0.02 & 21.3 & 21.5 \\ \hline
\multirow{2}{*}{GLM-130B} & 64 & 33540 & 0.06 & 22.4 & 52.4 & \multirow{2}{*}{GLM-130B} & 1024 & 11976 & 0.03 & 16.7 & 18.2 \\ \cline{2-6} \cline{8-12} 
 & 256 & 18776 & 0.04 & 17.2 & 19.4 &  & 4096 & 6040 & 0.02 & 19.2 & 20.1 \\ \hline\hline
\end{tabular}%
\caption{
    The search space and the time cost for \sysname on Heterogeneous GPUs.
  For the pictures of time cost, the light color without hatches represents the time spent searching, while the deep color with hatches represents the time spent simulating.
  We can observe that it only takes \sysname\ about 1 minute to complete the end-to-end simulation. 
}
\label{tab:exp:cost}
\end{table*}

\section{Experiments}\label{sec:exp}

To prove \sysname's optimal search ability on MegatronLM, we did a comparative analysis between \sysname\ and experts on MegatronLM in \S\ref{sec:exp:expert}.
Finally, we evaluate \sysname to search for the finance-optimal plan under different settings in \S\ref{sec:exp:finance}.

\begin{figure*}[thbp]
  \centering
    \subfloat{\includegraphics[width=0.4\textwidth]{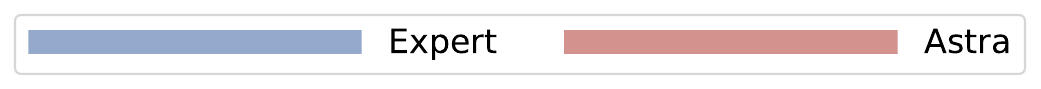}}\\
    \addtocounter{subfigure}{-1}

    \begin{minipage}{\textwidth}
    {\centering{\hspace{2.8cm}A800\hspace{4cm}H100\hspace{4.2cm}H800}}
    \end{minipage}

    \raisebox{0.8cm}{\rotatebox[origin=c]{90}{Llama-2}}
    \subfloat[7B]{\includegraphics[width=0.106\textwidth]{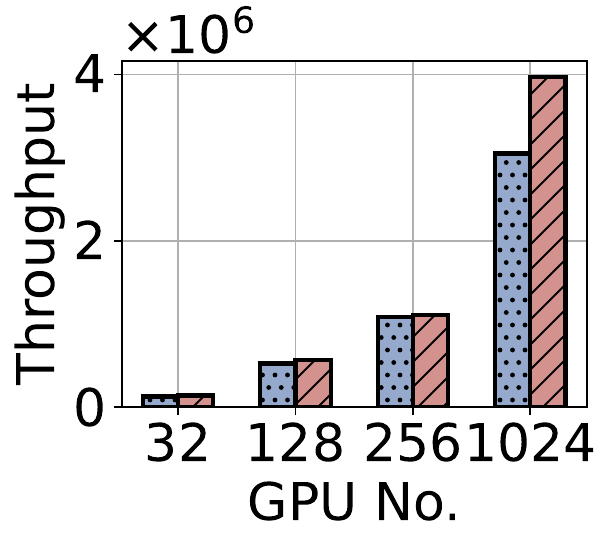}}
    \subfloat[13B]{\includegraphics[width=0.106\textwidth]{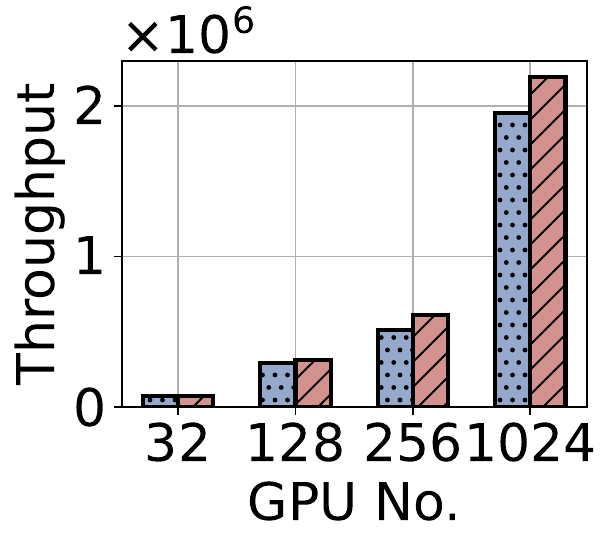}}
    \subfloat[70B]{\includegraphics[width=0.106\textwidth]{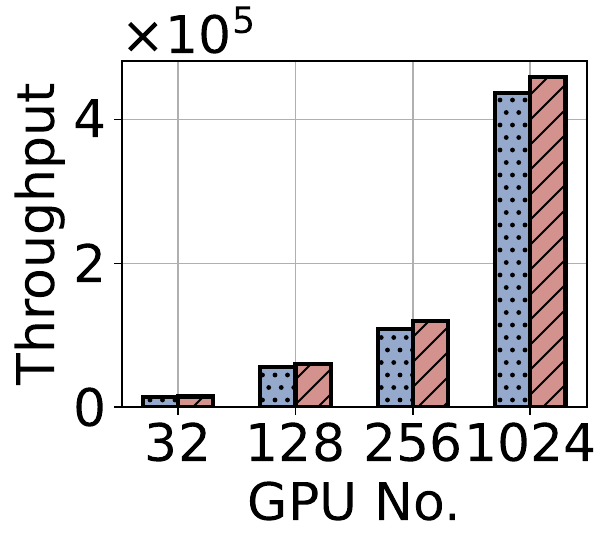}}
    \subfloat[7B]{\includegraphics[width=0.106\textwidth]{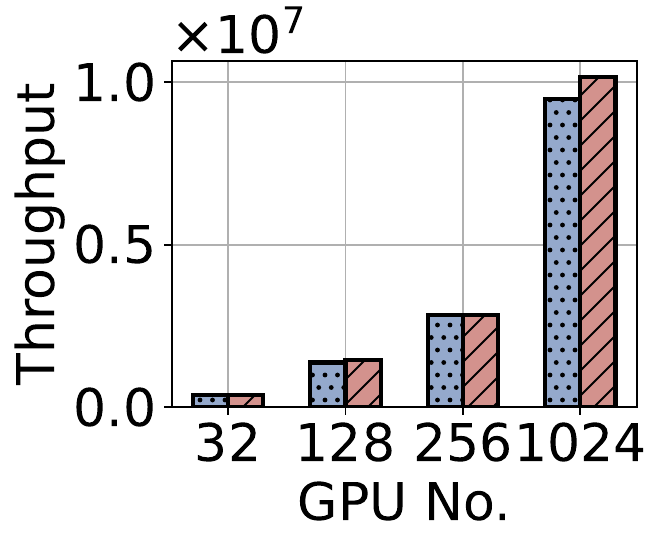}}
    \subfloat[13B]{\includegraphics[width=0.106\textwidth]{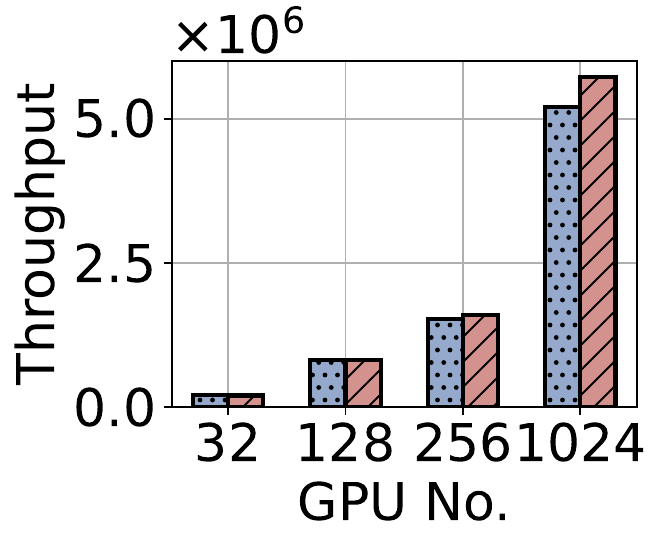}}
    \subfloat[70B]{\includegraphics[width=0.106\textwidth]{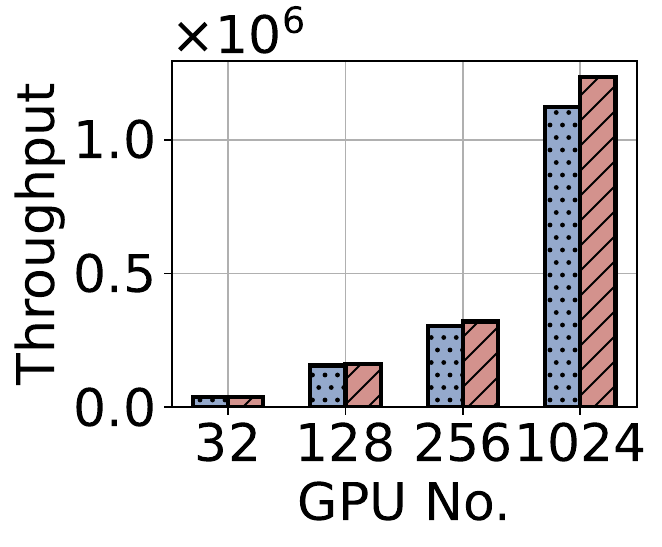}}
    \subfloat[7B]{\includegraphics[width=0.106\textwidth]{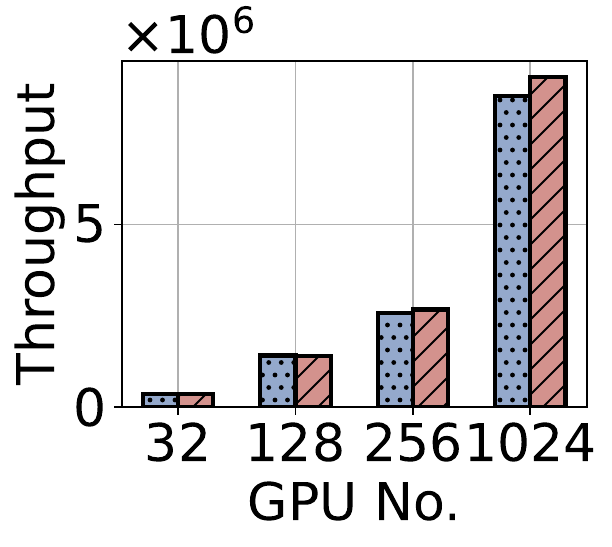}}
    \subfloat[13B]{\includegraphics[width=0.106\textwidth]{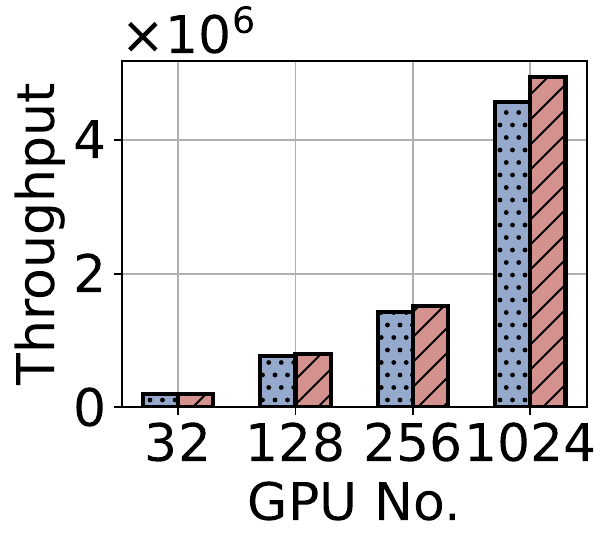}}
    \subfloat[70B]{\includegraphics[width=0.106\textwidth]{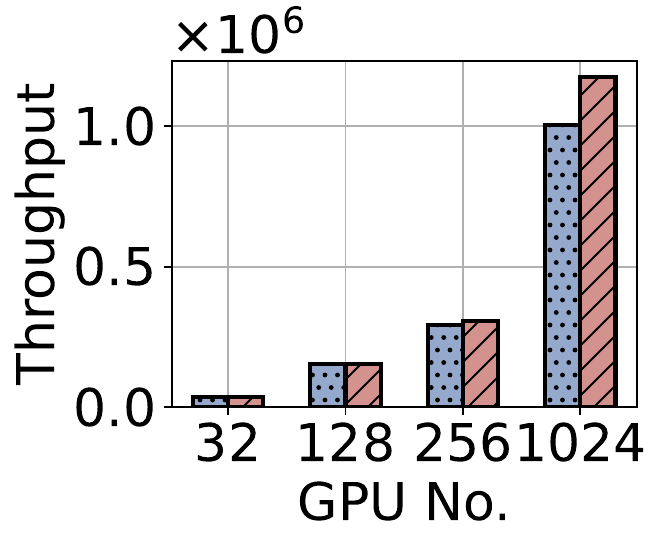}}
    \\
    \raisebox{0.8cm}{\rotatebox[origin=c]{90}{Llama-3}}
    \subfloat[8B]{\includegraphics[width=0.16\textwidth]{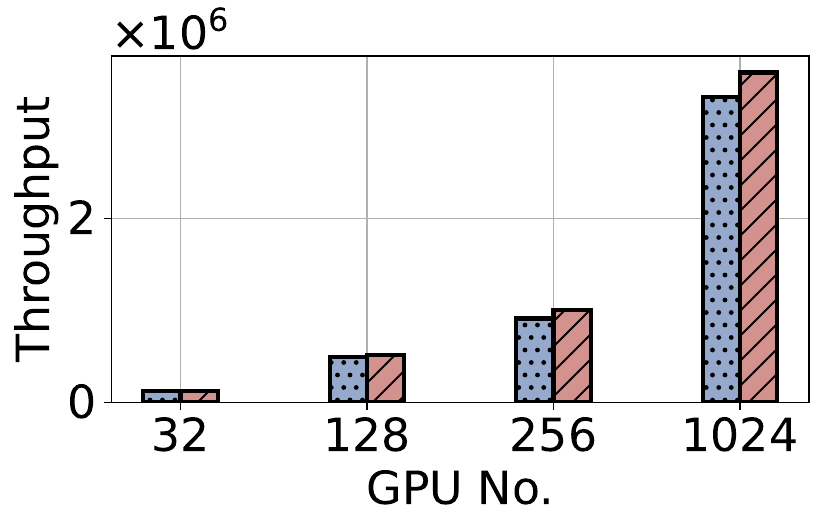}}
    \subfloat[70B]{\includegraphics[width=0.16\textwidth]{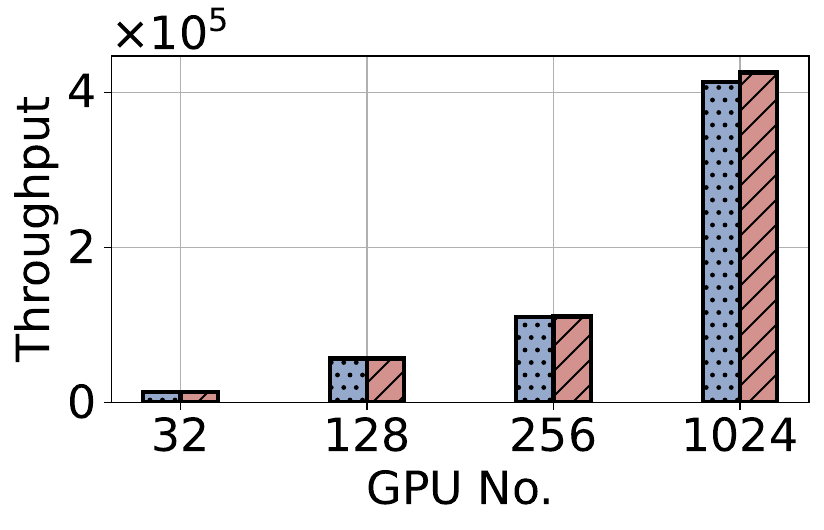}}
    \subfloat[8B]{\includegraphics[width=0.16\textwidth]{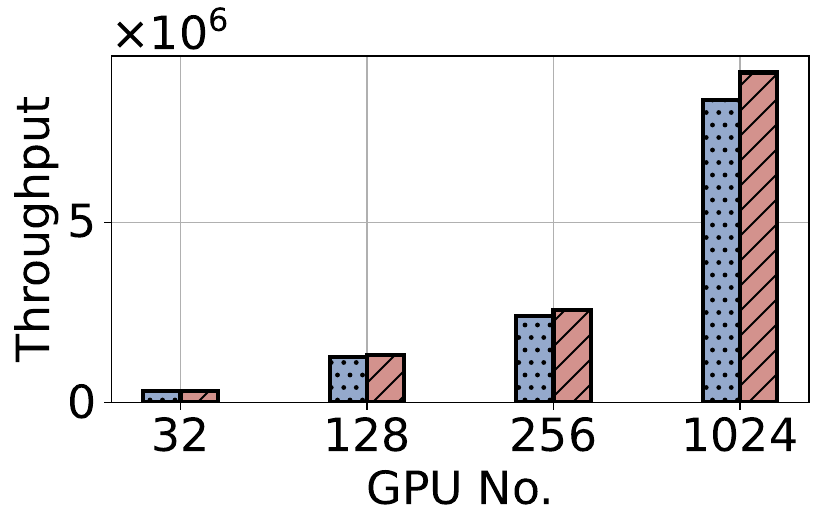}}
    \subfloat[70B]{\includegraphics[width=0.16\textwidth]{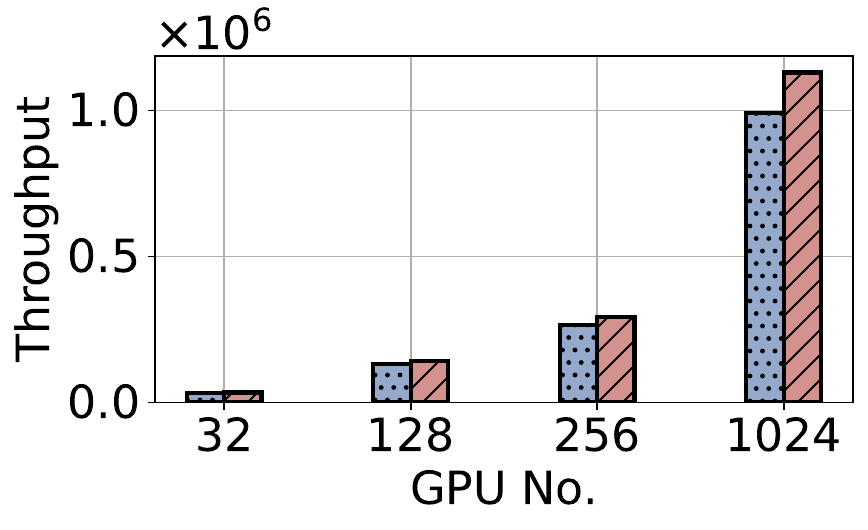}}
    \subfloat[8B]{\includegraphics[width=0.16\textwidth]{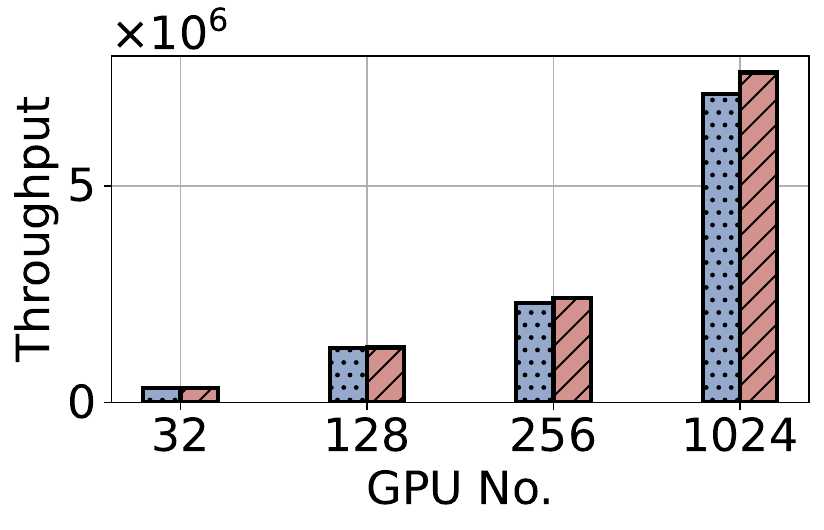}}
    \subfloat[70B]{\includegraphics[width=0.16\textwidth]{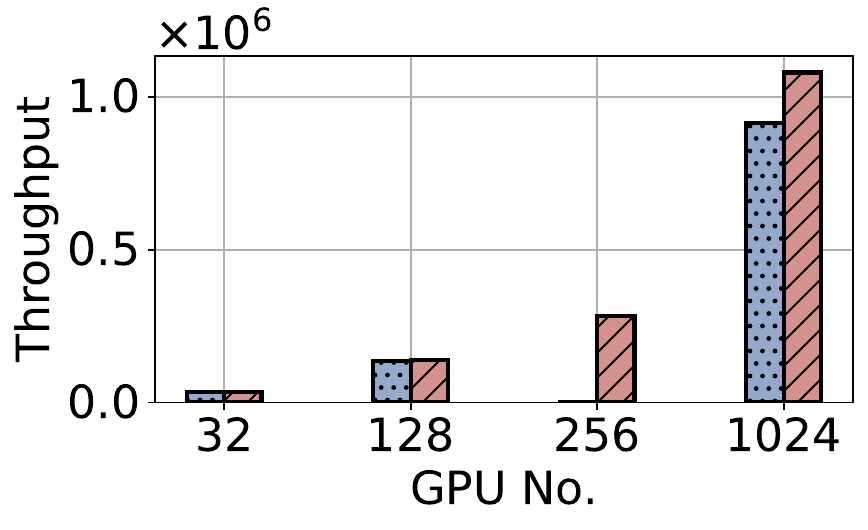}}
    \\
    \raisebox{0.8cm}{\rotatebox[origin=c]{90}{GLM}}
    \subfloat[67B]{\includegraphics[width=0.16\textwidth]{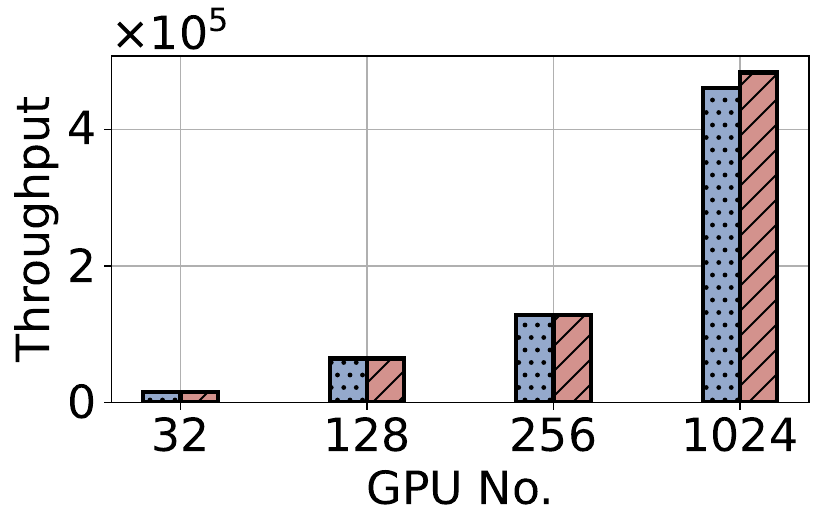}}
    \subfloat[130B]{\includegraphics[width=0.16\textwidth]{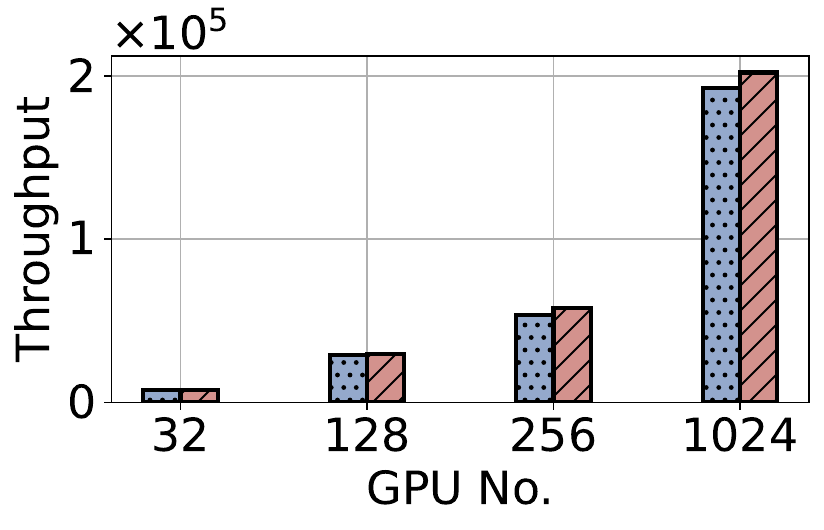}}
    \subfloat[67B]{\includegraphics[width=0.16\textwidth]{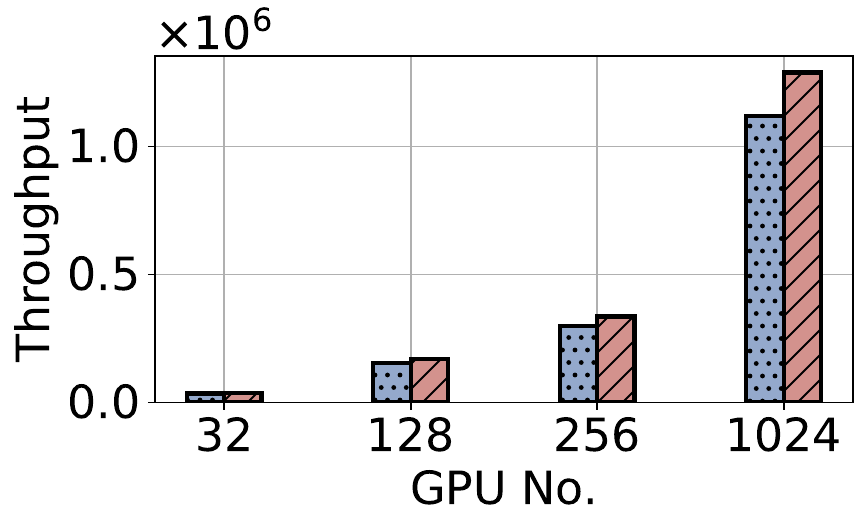}}
    \subfloat[130B]{\includegraphics[width=0.16\textwidth]{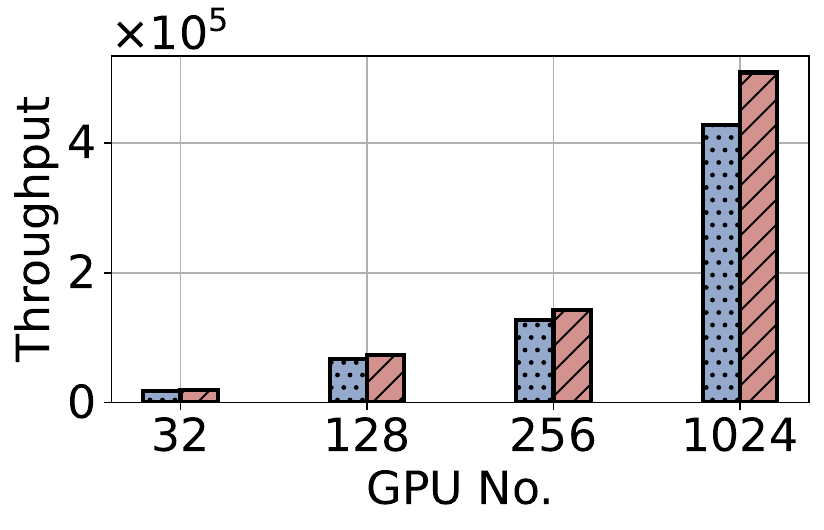}}
    \subfloat[67B]{\includegraphics[width=0.16\textwidth]{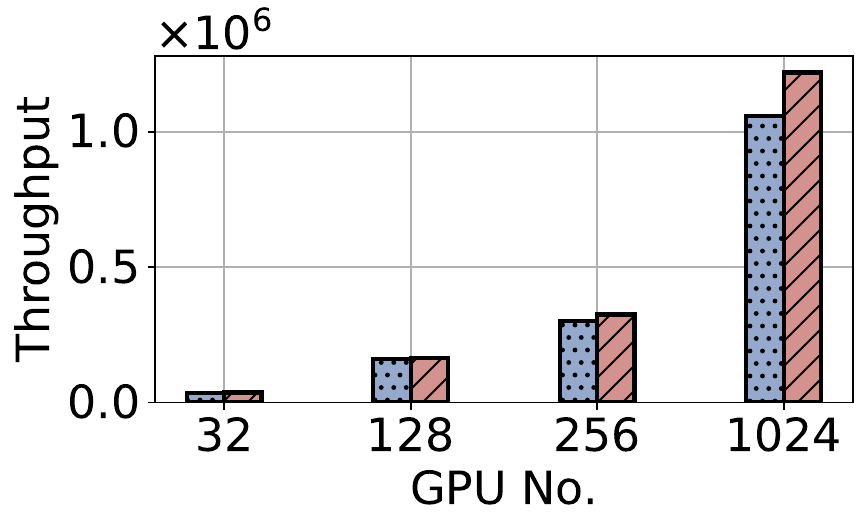}}
    \subfloat[130B]{\includegraphics[width=0.16\textwidth]{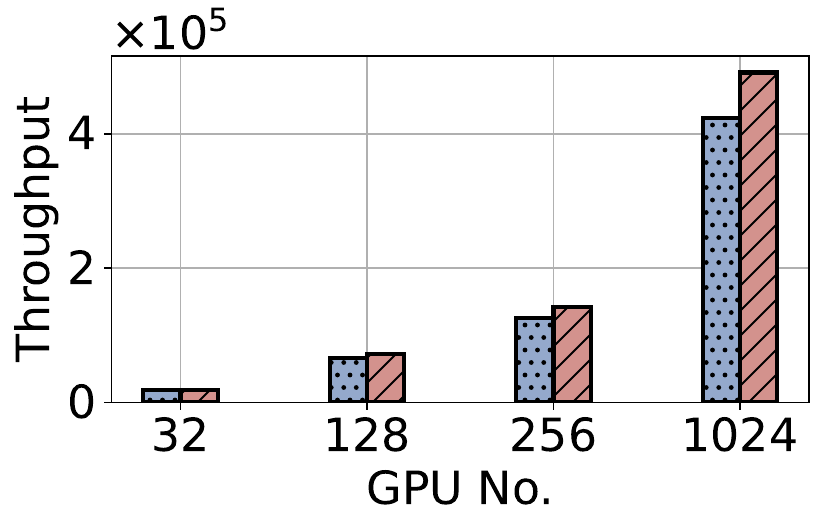}}
  \caption{
  We compare \sysname's searched optimal plan's throughput with expert's proposed plan's throughput in single-GPU setting.
  }
  \label{fig:expert:throughput}
  \vspace{-10pt}
\end{figure*}

\subsection{Mode-1: Comparison with Expert Plans}\label{sec:exp:expert}

\sssec{Method}.
To prove the \sysname's ability to search the optimal strategy on MegatronLM, we compared \sysname\ with an expert.
We first selected three models with different parameter sizes (7 model settings in total): Llama-2 (7B, 13B, and 70B), Llama-3 (8B, 70B), and GLM (67B, 130B).
Then, we offer 4 GPU number settings: 32, 128, 256, and 1024.
Next, we asked six experts to craft a parallel strategy for each setting (different models and different GPU settings, overall $7\times 4=28$ settings) based on their expert experience.
Each participant has over six years of industry machine learning service or training experience.
Then, we ran each of the six participants' parallel strategies for each setting on MegatronLM and picked the optimal one (one with the largest throughput) among the six expert-crated strategies as the expert-optimal strategy.
At last, we run \sysname\ to search the optimal parallel strategy automatically and compare the \sysname's parallel strategy's throughput with the expert-optimal parallel strategy's throughput.

\sssec{Results}.
As shown in Fig. \ref{fig:expert:throughput}, \sysname demonstrates its ability to automatically generate parallel strategies that match or exceed expert-tuned plans across various model configurations. This highlights \sysname's capability to generalize and optimize without manual intervention.

\par A key finding is that \sysname consistently matches or outperforms manually designed strategies, showing that its automated search can achieve results on par with domain experts. This adaptability extends across diverse hardware and model types, while specific setups often constrain expert-tuned plans. \sysname dynamically adjusts to different configurations, optimizing parallel strategies based on the specific training environment.

\par Another important observation is \sysname’s flexibility in combining different parallelism techniques—data, tensor, and pipeline. While expert strategies often focus on one type of parallelism, \sysname optimally balances multiple forms, leading to superior performance, especially for large-scale models. This hybrid approach is likely the key to future parallelism strategies, where flexibility and adaptation are critical.

\subsection{Mode-2: Heterogeneous GPU Search}

\begin{figure}[t]
  \centering
    \subfloat{\includegraphics[width=0.48\textwidth]{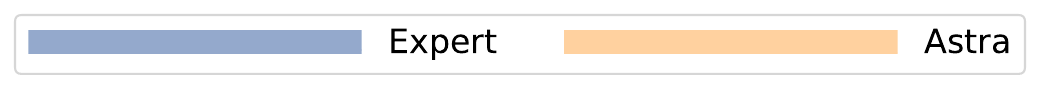}}\\
    \addtocounter{subfigure}{-1}
    
    \subfloat[Llama-2-7B]{\includegraphics[width=0.16\textwidth]{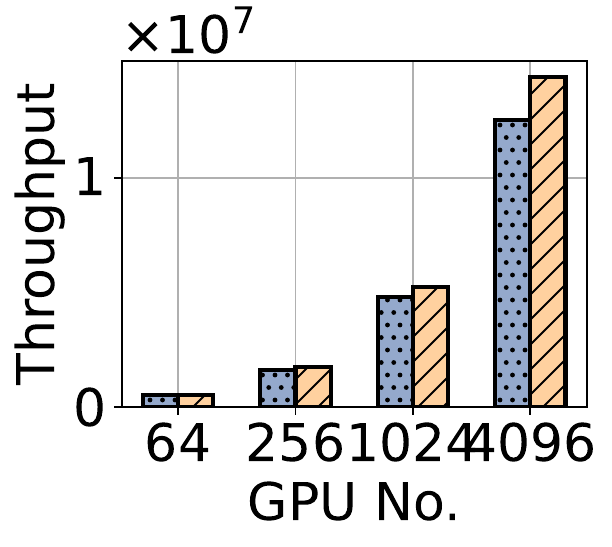}}
    \subfloat[Llama-2-13B]{\includegraphics[width=0.16\textwidth]{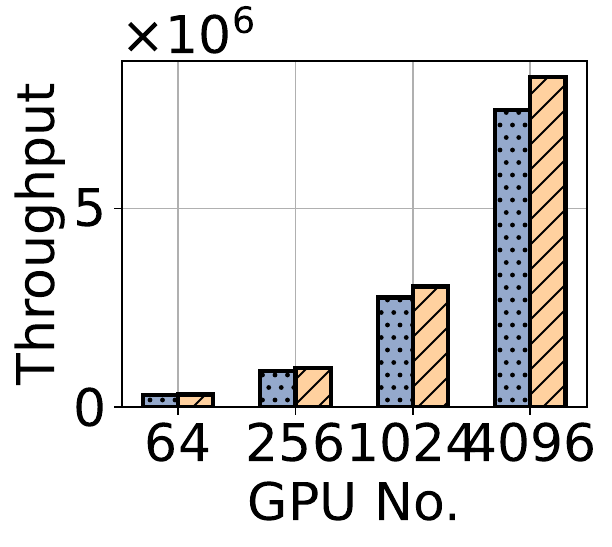}}
    \subfloat[Llama-2-70B]{\includegraphics[width=0.16\textwidth]{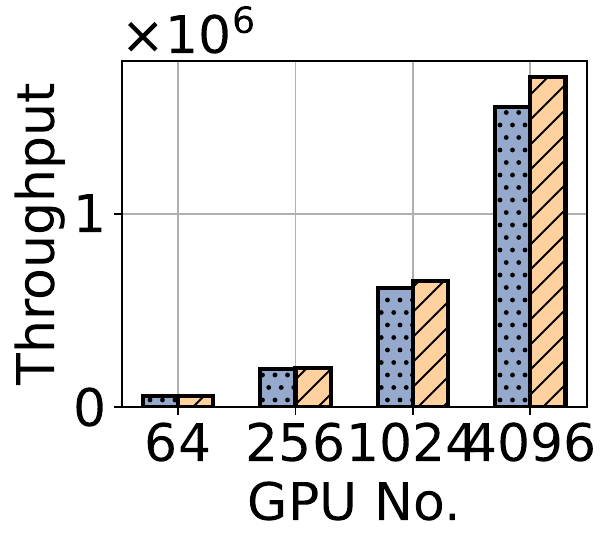}}
    \\

    \subfloat[Llama-3-8B]{\includegraphics[width=0.24\textwidth]{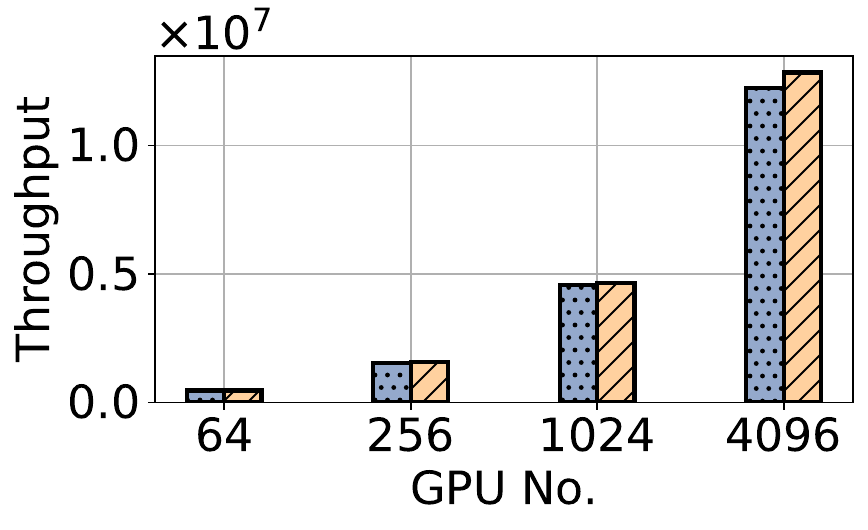}}
    \subfloat[Llama-3-70B]{\includegraphics[width=0.24\textwidth]{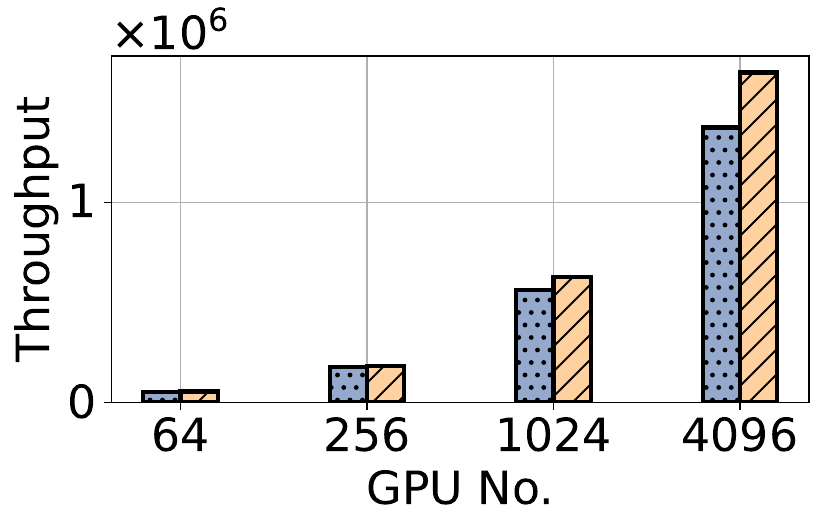}}
    \\

    \subfloat[GLM-67B]{\includegraphics[width=0.24\textwidth]{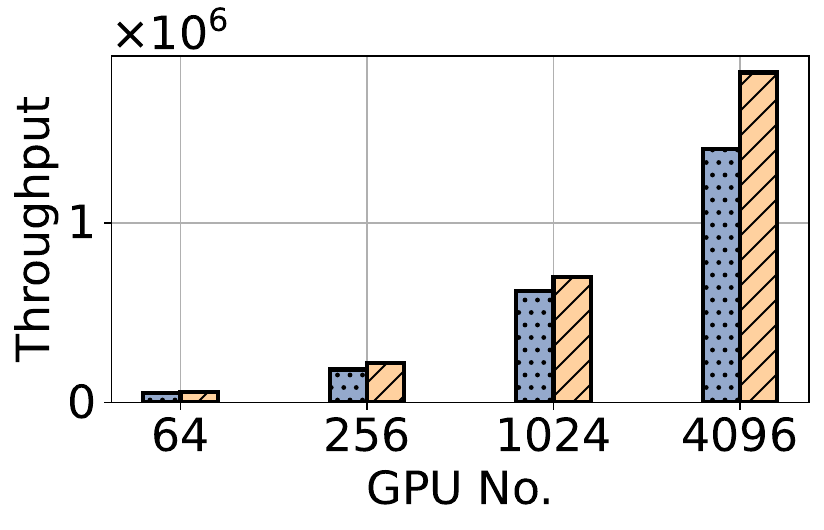}}
    \subfloat[GLM-130B]{\includegraphics[width=0.24\textwidth]{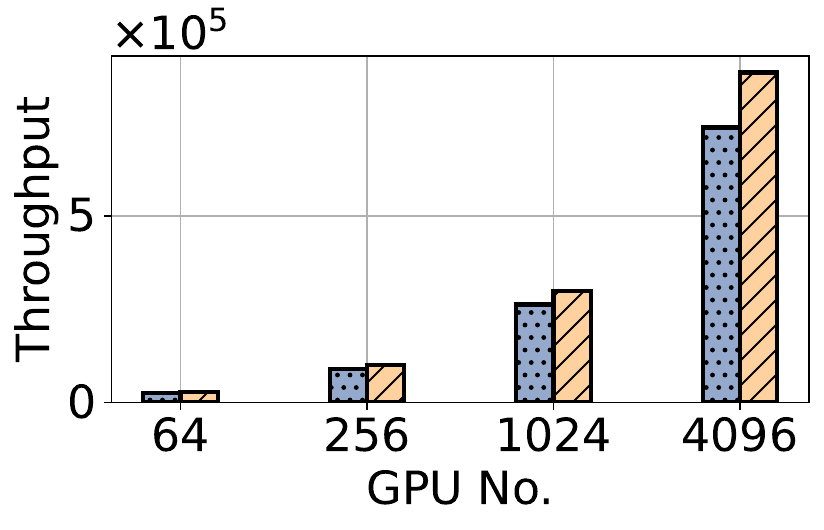}}
  \caption{
  For the heterogeneous GPU search scene, we compare expert-designed strategies's throughput with \sysname-searched strategies.
  The results prove the that \sysname achieves better throughput in heterogeneous scene.
  }
  \label{fig:exp:heter}
\end{figure}

\begin{table}[t]
\centering
\resizebox{0.5\textwidth}{!}{%
\begin{tabular}{c|cccc}
\hline
Model & H100 & H800 & A800 & Heter. \\ \hline\hline
Llama-2-7B & 10148287 & 9024716 & 3966756 & 5240609 \\
Llama-2-13B & 5721253 & 4937998 & 2187876 & 3040095 \\
Llama-2-70B & 1233850 & 1174362 & 458719 & 654206 \\
Llama-3-8B & 9167338 & 7610698 & 3586433 & 4660743 \\
Llama-3-70B & 1129568 & 1079507 & 425660 & 626050 \\
GLM-67B & 1288107 & 1218933 & 483384 & 699978 \\
GLM-130B & 508377 & 491088 & 202137 & 300193 \\ \hline\hline
\end{tabular}%
}
\caption{
We compare heterogeneous GPU with single-GPU search's optimal strategies' throughput.
The experiment is conducted with 1024 GPUs.
And the heterogeneous GPU setting is activated with A800 and H100.
}
\label{tab:exp:heter}
\end{table}

\sssec{Method}.
To evaluate \sysname's performance in heterogeneous GPU environments, we conducted a comprehensive comparison of \sysname-searched strategies and expert-designed strategies under heterogeneous GPU configurations. 
We use \sysname in the two GPU-heterogeneous environments with Nvidia H100 and A800 activated for search.
Also, we follow the design of \S\ref{sec:exp:expert}, we recruit six experts to craft a heterogeneous parallel strategy for each setting, and we picked the optimal one as the expert-designed strategy.
We offer 4 GPU number settings: 64, 256, 1024, and 4096.

Besides that, we also compared the heterogeneous GPU setting with single GPU setting in the same GPU number setting (1024).
We compare the throughput between the different settings (only A100, H100, H800, and heterogeneous settings)

\sssec{Results}.
As shown in Fig. \ref{fig:exp:heter}, our experiments reveal that \sysname consistently achieves higher throughput than expert-tuned configurations, particularly with larger models. \sysname’s approach dynamically balances data, tensor, and pipeline parallelism across heterogeneous GPUs, a task often challenging for manual tuning. This adaptability highlights the efficiency of automated strategies, especially in cloud-based or distributed environments where GPU types may vary. Overall, \sysname’s heterogeneous GPU search framework offers a scalable, cost-effective solution for optimizing model training in heterogeneous hardware contexts.

Table \ref{tab:exp:heter} shows the heterogeneous GPU setting compared with a single GPU setting.
Though a heterogeneous GPU setting strategy can not beat the performance of a single-GPU setting strategy, \sysname's searched strategy can nearly match with them.
\subsection{Mode-3: Evaluation Performance on Financial Cost}\label{sec:exp:finance}


\sssec{Search pools for GPU}. To comprehensively evaluate the financial cost performance of \sysname, we incorporate a variety of GPU types commonly used by major cloud service providers. Our search pools include the following GPU models: NVIDIA H100, A800 and H800.

These GPUs represent a range of performance capabilities and costs, providing a realistic and comprehensive basis for evaluating the financial efficiency of our system. By including these diverse GPU options, we can simulate the decision-making process of users who leverage cloud-based GPU resources, allowing us to optimize for both time and financial cost under various configurations.

\begin{figure}[t]
  \centering
    \subfloat[Per Throu. Llama-70B]{\includegraphics[width=0.24\textwidth]{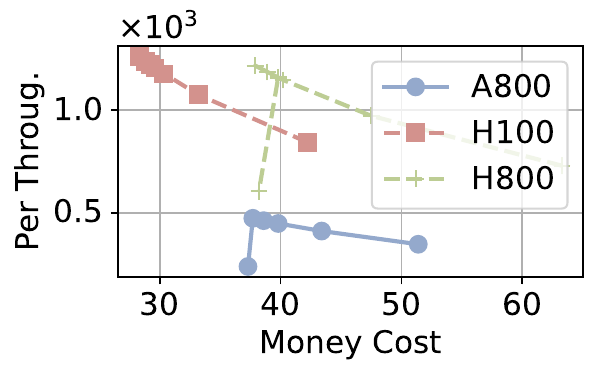}}
    \subfloat[Overall Throu. Llama-70B]{\includegraphics[width=0.24\textwidth]{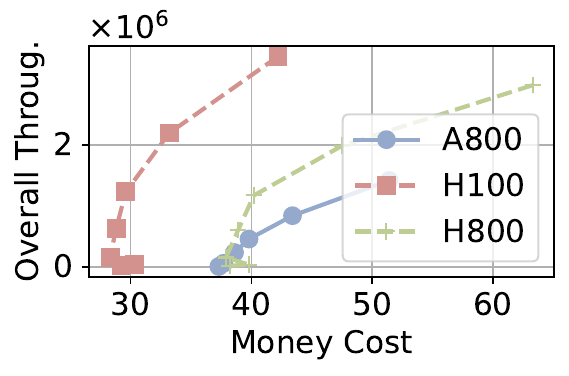}}
    \\
    \subfloat[Per Throu. GLM-67B]{\includegraphics[width=0.24\textwidth]{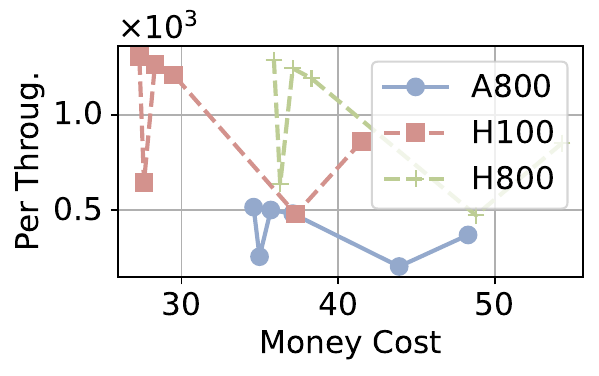}}
    \subfloat[Overall Throu. GLM-67B]{\includegraphics[width=0.24\textwidth]{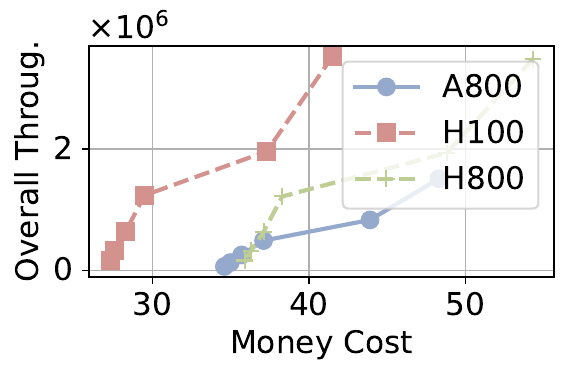}}
    \\
    \subfloat[Per Throu. GLM-130B]{\includegraphics[width=0.24\textwidth]{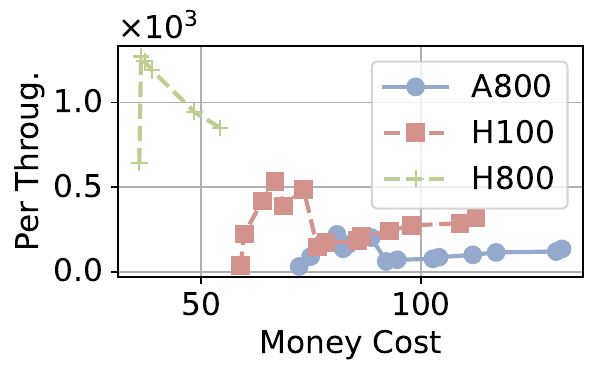}}
    \subfloat[Overall Throu. GLM-130B]{\includegraphics[width=0.24\textwidth]{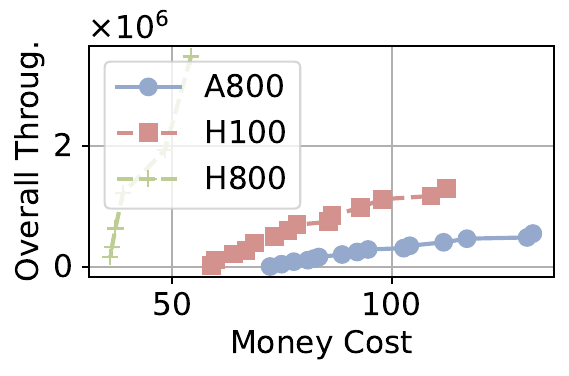}}
  \caption{
  We list the optimal line of \sysname.
  }
  \label{fig:money}
\end{figure}

\section{Discussion}

\sssec{Adaptation to heterogeneous environments}. \sysname's ability to adapt to dynamic and heterogeneous GPU clusters provides significant advantages for real-world deployments. In contrast to other frameworks that may be optimized for specific hardware setups, \sysname can adjust its strategies based on the available resources, making it particularly suited for cloud-based or distributed environments where hardware configurations may vary.

\sssec{Scalability and future directions}. The scalability demonstrated by \sysname, particularly in handling large models like Llama-70B, is a testament to its robustness. However, as models and GPU configurations continue to grow in complexity, further optimizations in the simulation phase could be crucial for maintaining performance. This suggests future work may focus on improving the efficiency of the simulation component to prevent bottlenecks as the search space expands.


\section{Conclusion}
\sysname proposes an automated parallelization strategy search framework that can optimize both training time and financial costs for the training of large-scale language models. In the experiment, \sysname was tested on the Llama series models (7B, 13B, 70B) and a variety of GPU numbers (32 to 4096). The results showed that its automatically generated parallel strategies were comparable to or even better than the strategies manually tuned by experts in terms of throughput. Especially under large-scale GPU clusters, \sysname's hybrid parallel strategy showed better scalability and efficiency. In addition, in the financial cost optimization experiment, \sysname was able to balance computing performance and cost through a variety of GPU configurations (such as H100, A800, etc.), providing targeted solutions for users with limited budgets. Overall, \sysname greatly reduces the complexity of parallel strategy tuning and provides strong support for efficient training of large-scale models in the future.





\clearpage
\bibliography{references}
\bibliographystyle{mlsys2025}

\clearpage
\newpage
\appendix

\section{Searchable Parameters}

\begin{table*}[htbp!]
    \centering
    \begin{tabular}{M{2.7cm}|M{5cm}|M{8cm}}
        \hline
        Parameter Category & Parallel Parameter & Parameter Range \\ \hline\hline
        \multirow{3}{*}{Parallel Strategy} 
        & pipeline-model-parallel-size & 1$\sim$N \\ \cline{2-3} 
        & tensor-model-parallel-size & 1$\sim$N \\ \cline{2-3} 
        & data-model-parallel-size & \$num\_gpus / \$(pipeline-model-parallel-size) / \$(tensor-model-parallel-size) \\ \hline
        \multirow{2}{*}{Parallelism}
        & micro-batch-size & 1$\sim$M \\ \cline{2-3} 
        & num-layers-per-virtual-pipeline-stage & 1$\sim$\$(pipeline-model-parallel-size) \\ \cline{2-3} 
        & sequence-parallel & [true, false] \\ \hline
        \multirow{1}{*}{Sharding Strategy}
        & use-distributed-optimizer & [true, false] \\ \hline
        \multirow{3}{*}{Recompute Strategy}
        & recompute-granularity & [selective, full, hybrid] \\ \cline{2-3} 
        & recompute-method & [block, uniform] \\ \cline{2-3} 
        & recompute-num-layers & 1$\sim$N \\ \hline
        \multirow{3}{*}{Offload Strategy}
        & offload-optimizer & [auto] \\ \cline{2-3} 
        & no-overlap-offload-optimizer & [true, false] \\ \hline
        \multirow{1}{*}{Computation Fusion}
        & use-flash-attn & [true] \\ \hline
        \multirow{4}{*}{Overlap Strategy}
        & overlap-p2p-communication & [true] \\ \cline{2-3} 
        & tp-comm-overlap & [true] \\ \cline{2-3} 
        & overlap-grad-reduce & [true] \\ \cline{2-3} 
        & overlap-param-gather & [true] \\ \hline
        \multirow{3}{*}{MoE Parameter}
        & num-experts & 1-N \\ \cline{2-3} 
        & expert-model-parallel-size & 1-N \\ \cline{2-3} 
        & moe-router-topk & 1-N \\ \hline\hline
    \end{tabular}
    \caption{
    The search parameters for \sysname\ to run on the MegatronLM backend.
    }
    \label{tab:parameter}
\end{table*}

We list the searchable parameters along with their range of values in Table \ref{tab:parameter}.

\section{Ablation Analysis}\label{sec:ablation}

In this section, we study the impact of different parallel techniques on performance.

\subsection{Experiment Setting}

We conducted experiments using the Llama-2 model in three sizes: 7B, 13B, and 70B, to evaluate the impact of various parallelism strategies and memory management techniques. The experiments were executed on GPU clusters composed of NVIDIA A800 GPUs, with configurations ranging from 64 to 4096 GPUs. We employed a hybrid parallelism strategy (data parallelism + tensor parallelism) unless otherwise specified. Four key techniques were analyzed: parallelism strategies, system scale, memory offloading, recomputation, and overlap strategies.

We tested the following setups for each technique:
\begin{itemize}[noitemsep,topsep=1pt, leftmargin=*]
    \item \textbf{Parallelism Strategy}: We tested data parallelism (DP), tensor parallelism (TP), pipeline parallelism (PP), and hybrid parallelism strategies across different GPU counts and model sizes.
    \item \textbf{System Scale}: We varied the number of GPUs (64, 128, 256, 1024, and 4096) while keeping the model architecture fixed to isolate the impact of system scale on training efficiency.
    \item \textbf{Memory Offloading}: We tested configurations with no offloading, enable offloading, and memory bandwidth variations (DDR4 vs. DDR5).
    \item \textbf{Overlap Strategy}: We tested no overlap, gradient communication overlap, parameter gather overlap, and combined overlap strategies.
\end{itemize}

\begin{figure*}[htbp]
  \centering
    \subfloat[Llama-2-7B]{\includegraphics[width=0.33\textwidth]{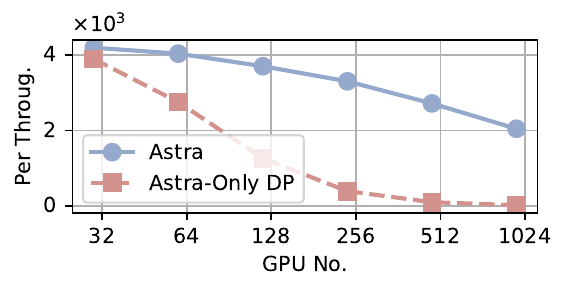}}
    \subfloat[Llama-2-13B]{\includegraphics[width=0.33\textwidth]{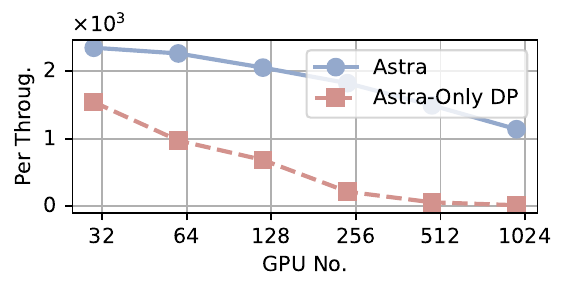}}
    \subfloat[Llama-3-8B]{\includegraphics[width=0.33\textwidth]{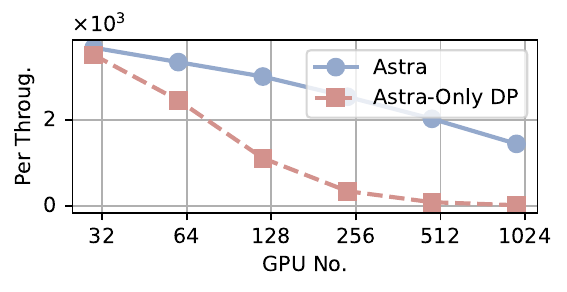}}
  \caption{
  We compare \sysname's performance with all parallelism methods allowed and only data parallelism allowed.
  }
  \label{fig:ablation:dp}
\end{figure*}

\begin{figure*}[htbp]
  \centering
    \subfloat[Llama-2]{\includegraphics[width=0.33\textwidth]{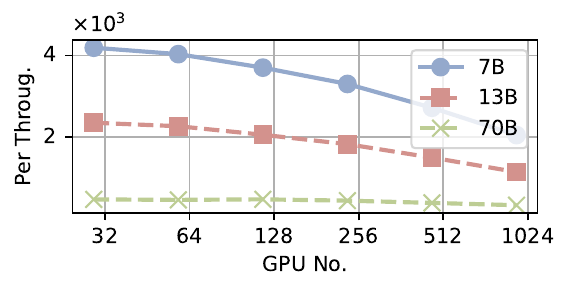}}
    \subfloat[Llama-3]{\includegraphics[width=0.33\textwidth]{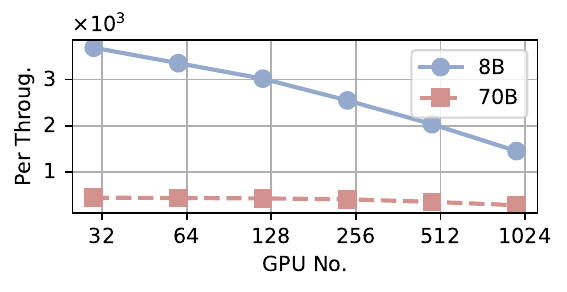}}
    \subfloat[GLM]{\includegraphics[width=0.33\textwidth]{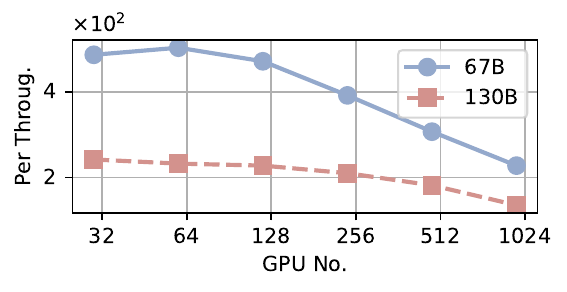}}
  \caption{
  We understand the system scale impact on training efficiency by training \sysname\ on different GPU numbers.
  }
  \label{fig:ablation:scale}
\end{figure*}

\sssec{Metrics}
We utilized three key metrics:
\begin{itemize}[noitemsep,topsep=1pt, leftmargin=*]
    \item \textbf{Training Throughput (samples/second)}: The primary metric for evaluating training efficiency, measuring the number of samples processed per second.
    \item \textbf{Scaling Efficiency (\%)}: This metric compares the achieved throughput at each scale to the ideal linear scaling expected as the number of GPUs increases.
    \item \textbf{Communication Overhead}: We measured the data transferred between GPUs to quantify the additional cost incurred by increasing the system size, focusing on inter-GPU and inter-node communication costs.
\end{itemize}

\sssec{Statistical Validation}
All experiments were run for a fixed number of iterations to ensure that the strategies had sufficient time to reach stable performance metrics. Each experiment was repeated three times to account for variability, and the results are reported as averages with standard deviations.

\subsection{Parallelism Strategy In-Depth Analysis}

\sssec{Method}. We analyzed the impact of different parallelism strategies (DP, TP, PP, and hybrid) on training performance across different GPU counts and model sizes. We monitored trade-offs between communication overhead and computation efficiency and identified cases where hybrid parallelism yielded improvements over single-strategy configurations.

\sssec{Results}.
Figure \ref{fig:ablation:dp} compares \sysname’s performance with all parallelism methods enabled against data parallelism (DP) alone across different system scales for Llama-2-7B, Llama-2-13B, and Llama-3-8B models. In all cases, \sysname consistently outperforms DP as the GPU count increases. The results show that \sysname maintains higher throughput by combining multiple parallelism strategies, effectively mitigating the communication overhead that causes DP's performance to degrade at larger scales. This demonstrates \sysname's superior scalability and adaptability, highlighting the benefits of hybrid parallelism over relying solely on data parallelism.

\subsection{System Scale Impact on Training Efficiency}

\sssec{Method}.
We explored the relationship between system scale (i.e., the number of GPUs) and training efficiency by analyzing changes in training throughput and communication overhead as the number of GPUs increased while keeping the model architecture fixed.

\sssec{Results}.
The results presented in Figure \ref{fig:ablation:scale} demonstrate a clear trend of diminishing per-GPU throughput as GPUs increase for each model (Llama-2, Llama-3, and GLM). For smaller models such as Llama-2 7B and 13B, the decrease in throughput per GPU is gradual, indicating that these models scale more efficiently with the number of GPUs. However, for larger models such as Llama-2 70B and GLM 130B, the throughput drops significantly as the GPU count exceeds 512, suggesting that these models experience higher communication overhead and resource contention at larger scales. 
This decline in efficiency highlights the limitations of scaling large models across many GPUs. As the system grows, the communication cost between GPUs begins to outweigh the computational benefits, leading to suboptimal utilization of the hardware resources. The sharp decrease in throughput for models like Llama-2 70B and GLM 130B at 1024 GPUs underscores the need for careful consideration of parallelization strategies to mitigate the impact of inter-GPU communication on training performance. These findings are consistent with previous observations in distributed training, where scaling efficiency degrades as the system size increases due to increased synchronization and data transfer costs.

\begin{figure}[htbp]
  \centering
    \subfloat{\includegraphics[width=0.4\textwidth]{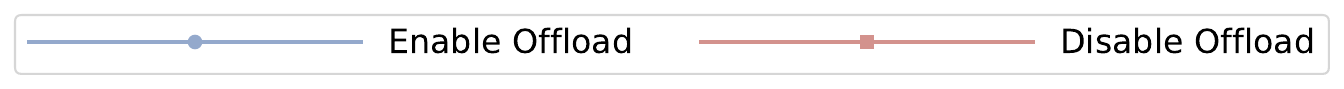}}\\
    \addtocounter{subfigure}{-1}
    
    \subfloat[Llama-2-7B]{\includegraphics[width=0.16\textwidth]{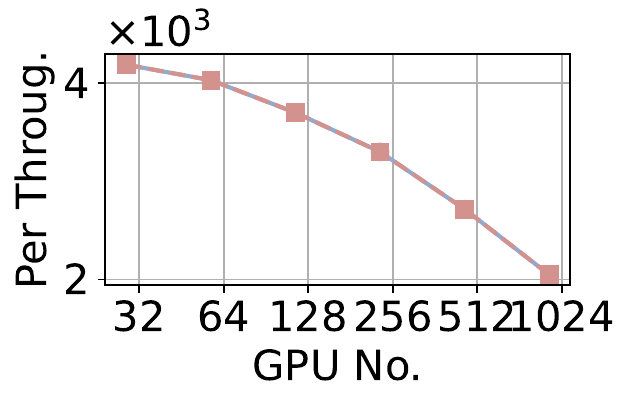}}
    \subfloat[Llama-2-13B]{\includegraphics[width=0.16\textwidth]{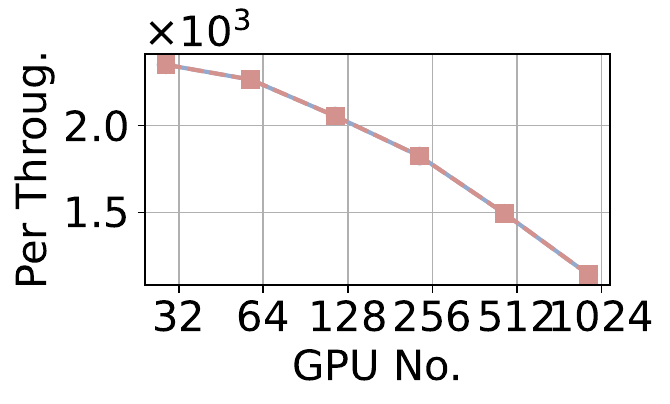}}
    \subfloat[Llama-2-70B]{\includegraphics[width=0.16\textwidth]{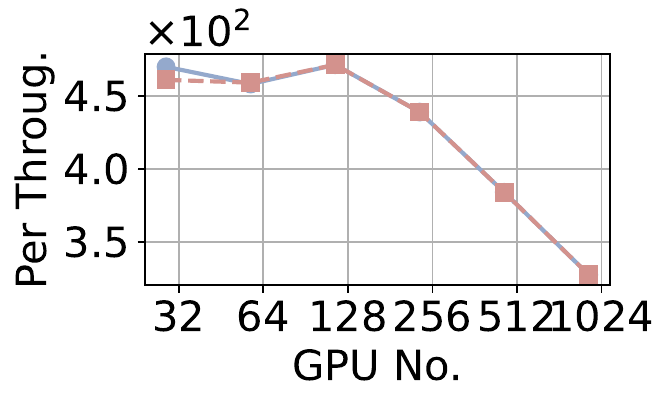}}
    \\
    \subfloat[Llama-3-8B]{\includegraphics[width=0.24\textwidth]{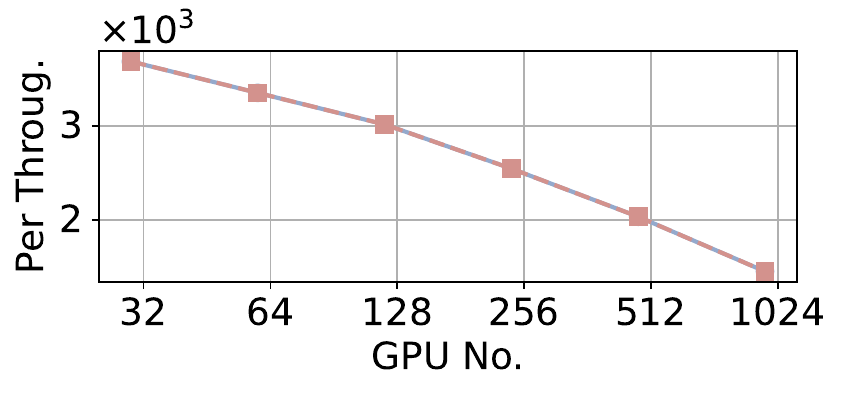}}
    \subfloat[Llama-3-70B]{\includegraphics[width=0.24\textwidth]{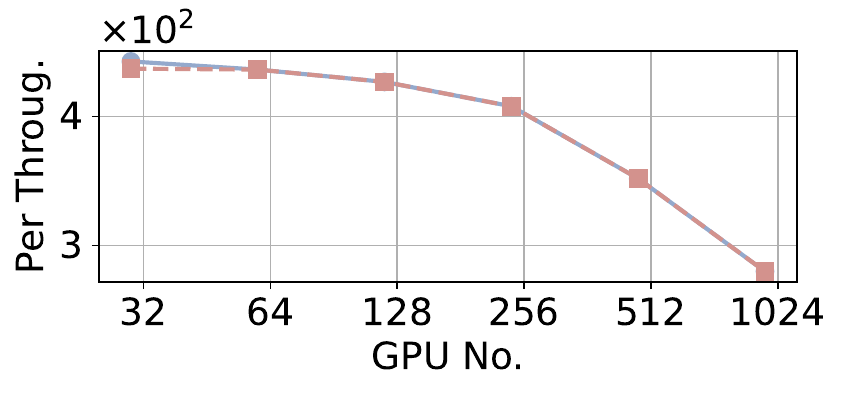}}
    \\
    \subfloat[GLM-67B]{\includegraphics[width=0.24\textwidth]{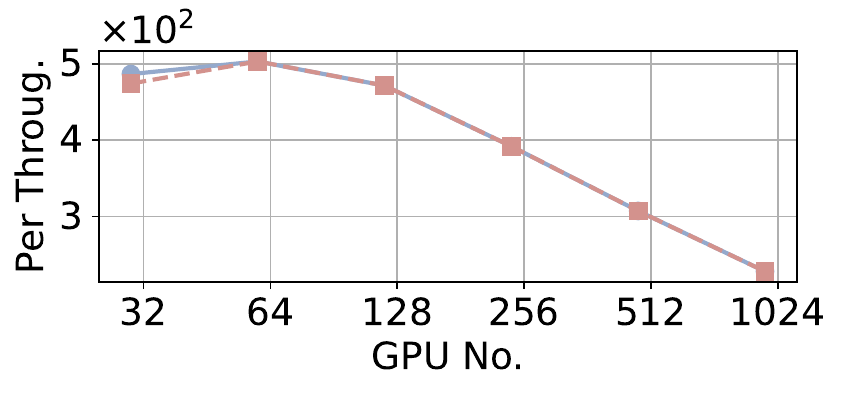}}
    \subfloat[GLM-130B]{\includegraphics[width=0.24\textwidth]{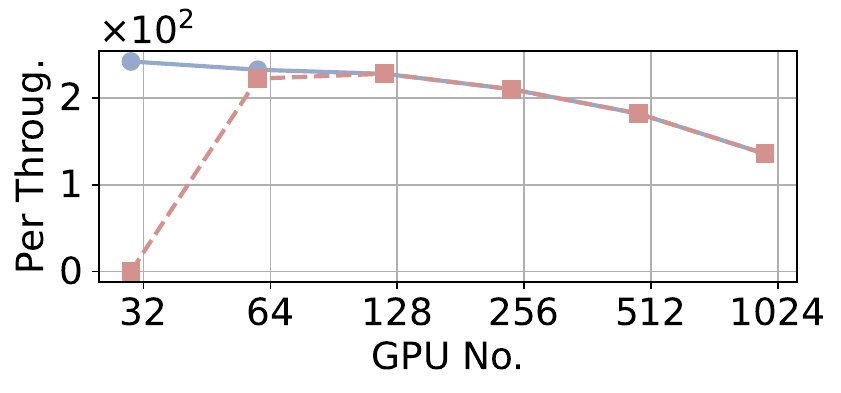}}
  \caption{
  We compare \sysname's performance with offload allowed with unallowed
  }
  \label{fig:ablation:offload}
\end{figure}

\subsection{Memory Offloading Technique Analysis}

\sssec{Method}. We evaluated the effectiveness of memory offloading techniques, comparing disable/enable offloading, and the impact of different memory bandwidths on training performance, particularly in memory-constrained environments.

\sssec{Results}.
Figure \ref{fig:ablation:offload} compares \sysname's performance with and without memory offloading across different models and system scales. The results show that memory offloading becomes increasingly important as model size grow. For smaller models like Llama-2-7B and Llama-2-13B, the performance impact of offloading is minimal, but as models scale up (e.g., Llama-70B and GLM-130B), enabling offloading significantly improves throughput by alleviating memory bottlenecks. Without offloading, larger models experience sharp performance declines as GPU count increases. These findings highlight the critical role of memory offloading in maintaining efficient scaling for larger models across large GPU configurations.




\begin{figure}[htbp]
  \centering
    \subfloat{\includegraphics[width=0.4\textwidth]{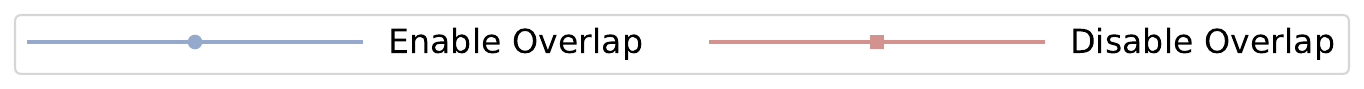}}\\
    \addtocounter{subfigure}{-1}
    
    \subfloat[Llama-2-7B]{\includegraphics[width=0.16\textwidth]{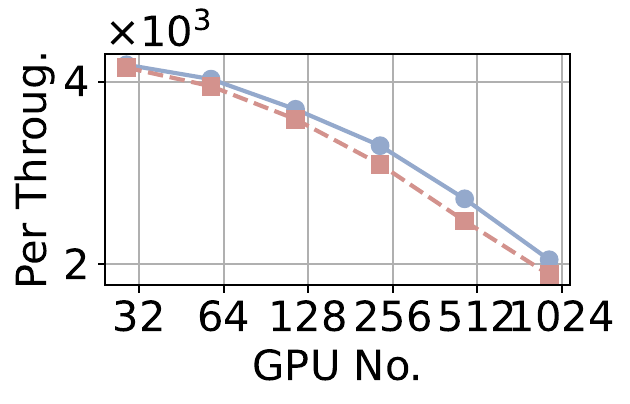}}
    \subfloat[Llama-2-13B]{\includegraphics[width=0.16\textwidth]{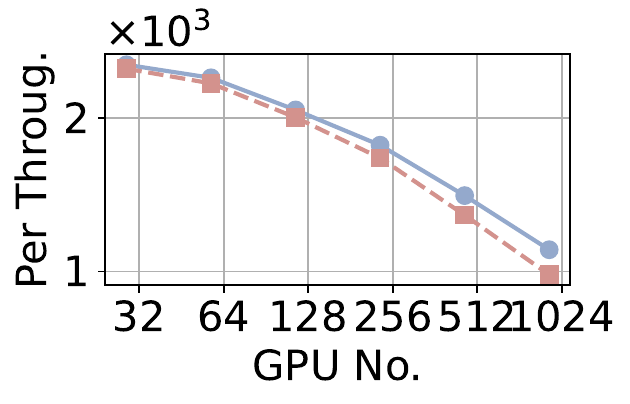}}
    \subfloat[Llama-2-70B]{\includegraphics[width=0.16\textwidth]{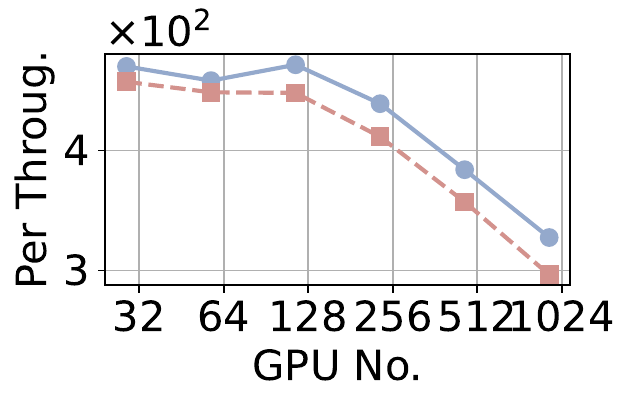}}
    \\
    \subfloat[Llama-3-8B]{\includegraphics[width=0.24\textwidth]{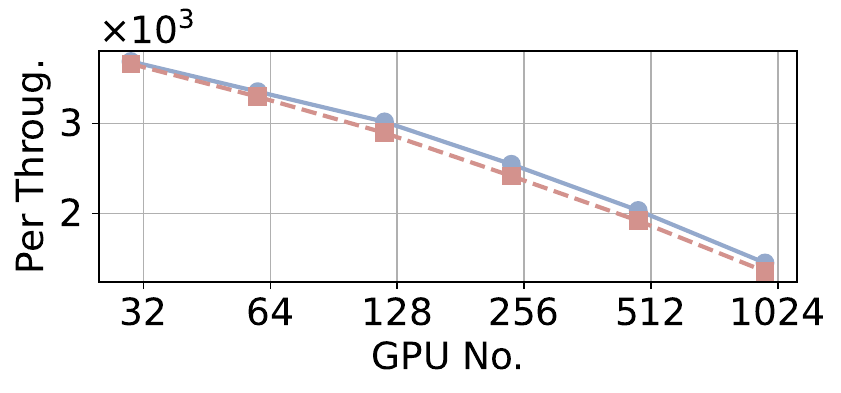}}
    \subfloat[Llama-3-70B]{\includegraphics[width=0.24\textwidth]{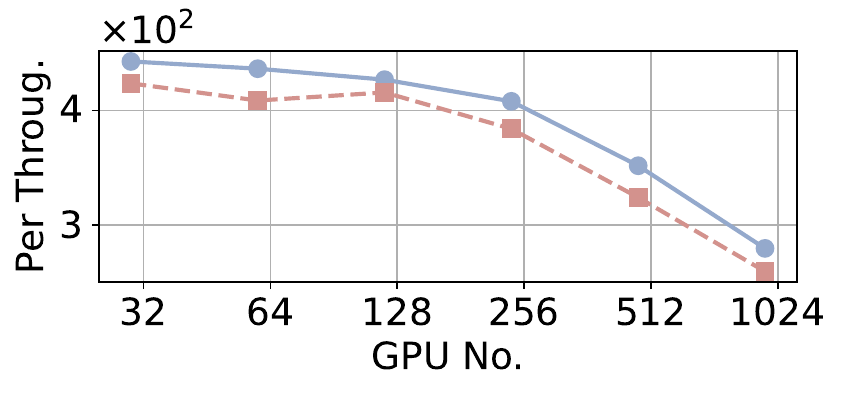}}
    \\
    \subfloat[GLM-67B]{\includegraphics[width=0.24\textwidth]{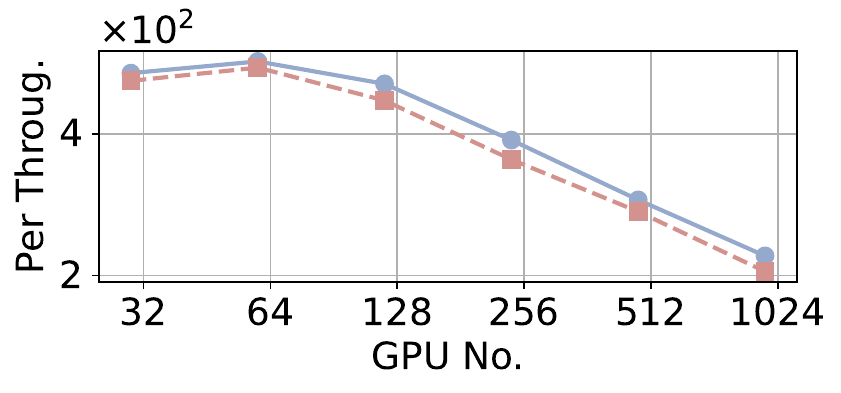}}
    \subfloat[GLM-130B]{\includegraphics[width=0.24\textwidth]{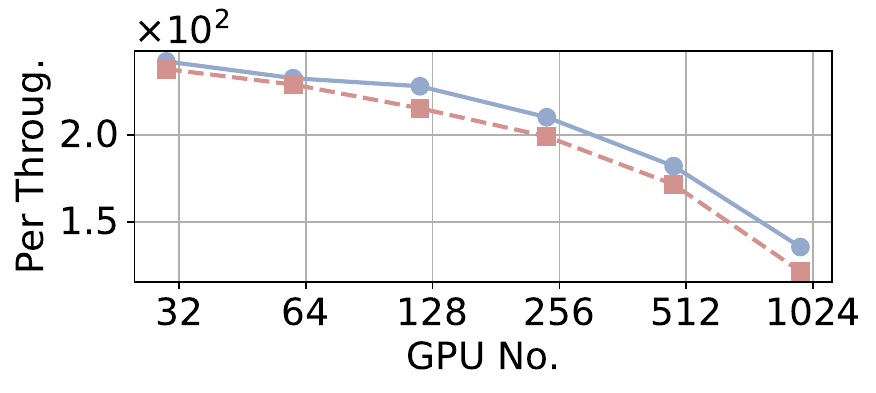}}
  \caption{
  We compare \sysname's performance with communication overlap allowed with unallowed
  }
  \label{fig:ablation:overlap}
\end{figure}

\subsection{Overlap Strategy Technique Analysis}

\sssec{Method}. We examined the effectiveness of overlapping communication with computation, comparing no overlap, gradient communication overlap, parameter gather overlap, and combined overlap strategies.

\sssec{Results}.
Figure \ref{fig:ablation:overlap} compares \sysname’s performance with and without communication overlap across different models and GPU scales. The results show that enabling communication overlap improves throughput, especially for larger models and higher GPU counts. For smaller models like Llama-2-7B and Llama-2-13B, the benefits are modest but noticeable, while for larger models such as Llama-2-70B and GLM-130B, overlap significantly reduces communication delays, resulting in better throughput. This highlights the importance of overlap strategies in optimizing performance and scalability, particularly for large-scale models where communication overhead becomes a bottleneck.

\section{Cost Analysis}\label{sec:exp:cost}

\sssec{Method}.
We did a cost analysis to show the gap between the large search space and the search efficiency of the \sysname.
We selected Llama-2 models (7B, 13B, and 70B) with 64, 256, 1024, and 4096 GPUs.
Then, for all the settings, we implemented \sysname\ on it and recorded the searched strategy number along with the end-to-end time (search time and simulation time)

\sssec{Result}. As shown in Table \ref{tab:exp:cost}, the number of explored strategies grows exponentially with model size. For smaller models like Llama-7B, even with 4096 GPUs, the search space remains relatively small. However, for larger models such as Llama-70B, the search space nearly triples compared to Llama-7B under the same GPU configuration. The end-to-end time reveals that the simulation phase is the main bottleneck, which may take 1 minute to execute on average. While the search time only takes less than 1 second to execute on average. This highlights the need for optimizing the simulation process, particularly in large-scale settings, while \sysname’s search algorithm remains efficient and scalable across different configurations.


\end{document}